\documentclass[a4paper,11pt]{article}
\usepackage{iftex}
\ifPDFTeX\pdfoutput=1\fi
\usepackage{jheppub}
\usepackage{amsmath,amssymb,mathtools,booktabs,array,graphicx,slashed,xcolor}
\usepackage{placeins}
\newcommand{\paperfigure}[1]{%
  \IfFileExists{figures/#1.pdf}{%
    \includegraphics[width=\textwidth]{figures/#1.pdf}%
  }{%
    \fbox{\parbox[c][4cm][c]{0.94\textwidth}{\centering
      Figure PDF\\[6pt]
      \texttt{\detokenize{figures/#1.pdf}}\\[6pt]
      The caption and figure reference are retained.}}%
  }%
}
\title{Saxion Decay and Dark Radiation in the KSVZ Axion Model}
\newcommand{\paperabstracttext}{%
We calculate the relativistic axion energy density produced by the decay of a curvature-displaced Peccei--Quinn condensate in a quartic KSVZ model. We follow the radial PQ field (the saxion) as it relaxes from its inflationary minimum and determine the surviving condensate abundance from the adiabatic action of its oscillations. We then evolve the condensate through its decay into axions and visible particles to calculate the resulting dark-radiation contribution, expressed as $\Delta N_{\rm eff}$. With one heavy coloured KSVZ fermion and no additional portal
interactions included, the axion channel dominates the two
tree-level decay modes. Two reference histories yield $\Delta N_{\rm eff}\simeq0.033$ and $0.12$, with maximum saxion energy fractions below $1\%$ and $4\%$, respectively. At fixed quartic coupling, increasing the duration of curvature release reduces the predicted axion radiation density relative to visible radiation by a factor of approximately $28$ while preserving the primordial isocurvature amplitude.
}

  \author[a]{Sidhartha Samtani}
  \author[b,c]{Amartya Sengupta}
  \affiliation[a]{Homer L. Dodge Department of Physics and Astronomy, University of Oklahoma,
    Norman, OK 73019, USA}
  \affiliation[b]{Department of Physics, The State University of New York, Buffalo, New York 14260,
    USA}
  \affiliation[c]{Fermi National Accelerator Laboratory, Batavia, Illinois 60510, USA}
  \emailAdd{sidhartha.samtani@ou.edu}
  \emailAdd{amartyas@buffalo.edu}
  \abstract{\paperabstracttext}

\begin{document}
\maketitle

\section{Introduction}
\label{sec:introduction}

The Peccei--Quinn (PQ) mechanism provides a dynamical solution to the strong CP problem and introduces the QCD axion~\cite{Peccei:1977hh,Peccei:1977ur,Weinberg:1977ma,Wilczek:1977pj,Kim:2008hd,Hook:2018dlk,Agrawal:2022yvu}.\footnote{Reviews of astrophysical axion bounds and the broader phenomenology of meV axions appear in Refs.~\cite{Caputo:2024oqc,Cicoli:2026fqp}.} The axion can account for dark matter through vacuum realignment~\cite{Preskill:1982cy,Abbott:1982af,Dine:1982ah,Marsh:2015xka,OHare:2024nmr,Arias:2012az}. If PQ symmetry breaks before inflation and is not subsequently restored, the observable Universe inherits an approximately homogeneous axion angle~\cite{Safdi:2022xkm}. Quantum fluctuations during inflation nevertheless perturb this angle and later generate cold-dark-matter isocurvature, which is tightly constrained by cosmic microwave background (CMB) observations~\cite{Planck:2018jri,Petretti:2026ayw}.

The resulting constraint depends on the PQ radius during inflation. A radius larger than its vacuum value suppresses angular fluctuations because the same canonical field displacement corresponds to a smaller change in angle~\cite{Linde:1991km,Kearney:2016vqw,Kobayashi:2016qld,Kawasaki:2008jc}. A coupling to spacetime curvature can produce such a displacement, with a small quartic coupling allowing a large inflationary radius. As the curvature decreases after inflation, the radial minimum moves towards the vacuum. The field can then develop large oscillations that amplify axion fluctuations through tachyonic growth and parametric resonance~\cite{Kasuya:1996ns,Harigaya:2015hha,Ema:2017krp,Ballesteros:2021bee}. The viability of inflationary isocurvature suppression therefore also depends on the subsequent radial evolution.

Graham and Racco~\cite{Graham:2025iwx} demonstrated that a quartic PQ theory admits curvature-displaced histories in which fluctuation growth remains sufficiently small, and showed that the outcome depends on the reheating history. The energy remaining in radial oscillations determines the saxion abundance after relaxation. Once the radial oscillations become approximately harmonic, their cycle-averaged energy density redshifts as nonrelativistic matter. Its density decreases more slowly than radiation, so an initially small condensate can become cosmologically relevant before it decays. The relativistic axions produced by saxion decay increase the effective number of neutrino species by $\Delta N_{\rm eff}$, providing an additional constraint on PQ cosmology~\cite{Moroi:2013tea,Moroi:2014mqa,Dror:2021nyr,Chang:1996ih,Ichikawa:2007jv,Graf:2012hb}.

Primordial isocurvature depends on the inflationary axion phase power, which is proportional to $[H_*/f_R(t_*)]^2$. Here $t_*$ denotes the time when the CMB reference scale leaves the Hubble radius, $H_*=H(t_*)$ is the Hubble rate at that time, and $f_R(t_*)$ is the PQ radius at the inflationary minimum. The subsequent dark-radiation abundance depends on the energy remaining in radial oscillations and its conversion into axions through saxion decay. CMB measurements of the radiation density can therefore test histories that satisfy the primordial isocurvature bound~\cite{AtacamaCosmologyTelescope:2025nti,Planck:2018vyg}. Connecting these observables requires following the radial field through relaxation to determine the surviving saxion abundance, since a large initial displacement alone does not specify how much oscillation energy remains when the minimum reaches its vacuum value.

Axion dark radiation from displaced PQ condensates has been studied in several previous works~\cite{Co:2016xti,Co:2017orl,Chun:2000jr,Jeong:2012np,Bae:2013qr}. Mazumdar, Qutub and Saikawa~\cite{Mazumdar:2016nzr} considered a non-supersymmetric quartic PQ field displaced during inflation. Using analytic estimates of its evolution and an instantaneous-decay approximation, they calculated the axion radiation and constrained the PQ parameters and reheating temperature, including cases where the condensate remains subdominant.

The surviving condensate also depends on how it relaxes and exchanges energy with other fields. Previous studies have examined adiabatic suppression of saxion oscillations~\cite{Kawasaki:2011aa}, thermal trapping and dissipation~\cite{Moroi:2013tea,Moroi:2014mqa}, and nonperturbative axion production from large radial oscillations~\cite{Ema:2017krp}. Koz{\'o}w and Olechowski~\cite{Kozow:2022whq} further studied how radiative, thermal and spacetime-curvature corrections affect PQ evolution, axion production and isocurvature at very small self-coupling.

Axion radiation has also been calculated from derivative inflaton couplings~\cite{Kaneta:2024yyn} and from inflaton decay and annihilation in a shift-symmetric effective theory~\cite{Xu:2026wnv}. Here we follow a spectator PQ condensate through curvature release and determine its surviving oscillation action, which determines the abundance available for late saxion decay. Varying the duration of curvature release at fixed primordial angular normalization and particle-physics parameters allows us to isolate how reheating changes the resulting $\Delta N_{\rm eff}$.

We perform this calculation in the Kim--Shifman--Vainshtein--Zakharov (KSVZ) model with one coloured Dirac fermion~\cite{Kim:1979if,Shifman:1979if,DiLuzio:2017pfr}. Its Yukawa interaction determines the fermion mass and saxion decay threshold, as well as the thermal force on the PQ field. We assume a negligible PQ--Higgs portal and work in a weak-Yukawa regime where early relaxation and late perturbative decay occur on well-separated timescales. We assess thermal forces, dissipation and thermal noise using the framework of Refs.~\cite{Mukaida:2012qn,Moroi:2013tea,Moroi:2014mqa} and adopt the decay amplitudes of Ref.~\cite{Benabou:2023ghl}.

We follow a range of smooth transitions to radiation domination, evolve the radial background and its fluctuations, and extract the surviving oscillation action after the field settles near a broken minimum. We calculate the initial condensate abundance from this action and evolve
the saxion, visible bath and relativistic axions as a coupled system. The reference calculations show that saxion decay can generate appreciable dark radiation even when the condensate contributes only a few percent of the total energy density. At fixed quartic coupling, a more gradual curvature release leaves less coherent energy available for decay, lowering $\Delta N_{\rm eff}$ by nearly a factor of $28$ across the sampled histories without changing the primordial angular power.

Sections~\ref{sec:model} and \ref{sec:decay} introduce the model and its decay regime. Sections~\ref{sec:numerics} and \ref{sec:results} present radial relaxation and the resulting radiation abundance. Section~\ref{sec:discussion} discusses the interpretation and limitations of the calculation, and Section~\ref{sec:conclusions} summarizes the results and future directions. The appendices contain the particle-physics details, thermal calculations and numerical checks.

\section{The axion model}
\label{sec:model}

We consider a complex scalar $\Phi$ whose vacuum expectation value spontaneously breaks the Peccei--Quinn symmetry. A positive coupling to spacetime curvature displaces the field to a larger radius during inflation. As the curvature decreases, the minimum returns towards its vacuum value, exciting the radial motion that supplies the saxion condensate.

We work in a spatially flat Universe with scale factor $a(t)$, Hubble rate $H=\dot a/a$ and metric signature $(+---)$, where a dot denotes differentiation with respect to cosmic time. Our curvature convention is $R=6(2H^2+\dot H)$, and the reduced Planck mass is $M_{\rm P}=2.435\times10^{18}\,\mathrm{GeV}$. For a scalar of unit PQ charge, the relevant Lagrangian is
\begin{equation}
 \mathcal L_{\rm PQ}=|\partial\Phi|^2
 -\lambda_\Phi\left(|\Phi|^2-\frac{f_a^2}{2}\right)^2
 +2\xi R|\Phi|^2
 +(\theta+\theta_{\rm QCD})
 \frac{\alpha_s}{8\pi}G^A_{\mu\nu}\widetilde G_A^{\mu\nu}.
 \label{eq:model_Lagrangian}
\end{equation}
Here $\lambda_\Phi>0$ is the quartic coupling, $\xi>0$ is the nonminimal curvature coupling, and $\theta=\arg\Phi$. The last term couples the axion angle to the gluon field strength $G^A_{\mu\nu}$ and its dual $\widetilde G_A^{\mu\nu}$, with $\alpha_s$ the strong coupling and $\theta_{\rm QCD}$ the constant strong CP angle. This term represents the low-energy anomaly induced by the KSVZ fermion~\cite{GEORGI198673,Bauer:2020jbp,Reece:2023czb}. We take domain-wall number one, so the vacuum PQ radius $f_a$ also coincides with the axion decay constant~\cite{DiLuzio:2017pfr}. The QCD potential is negligible during the radial relaxation considered here~\cite{GrillidiCortona:2015jxo}.

We use polar fields to describe radial and angular motion, and Cartesian fields to follow the trajectory through the origin. We use
\begin{equation}
 \Phi=\frac{\rho}{\sqrt2}e^{i\theta}
 =\frac{X+iY}{\sqrt2},
 \qquad
 \rho^2=X^2+Y^2,
 \qquad
 a_{\rm ax}=f_a\theta,
 \label{eq:model_fields}
\end{equation}
where $a_{\rm ax}$ is the canonically normalized axion near the vacuum. In Cartesian variables, the scalar potential is
\begin{equation}
 V_0=\frac{\lambda_\Phi}{4}(X^2+Y^2-f_a^2)^2
 -\xi R(X^2+Y^2).
 \label{eq:model_potential}
\end{equation}
The curvature term lowers the quadratic coefficient and moves the broken minimum outwards. Minimizing the potential with respect to the radius determines its instantaneous location,
\begin{equation}
 f_R^2=f_a^2+\frac{2\xi R}{\lambda_\Phi}.
 \label{eq:model_fR2}
\end{equation}
During inflation, $R\simeq12H^2$, so a sufficiently small quartic coupling allows $f_R\gg f_a$. The radial mass at this minimum is $m_{s,\min}=\sqrt{2\lambda_\Phi}f_R$. As the curvature vanishes, the minimum approaches $f_a$ and the mass approaches the vacuum saxion mass $m_s=\sqrt{2\lambda_\Phi}f_a$.

We denote homogeneous fields by an overbar and choose the initial field direction along the $X$ axis, with vanishing angular velocity, so $\overline Y=0$. The orientation relative to the eventual QCD minimum remains encoded in $\theta_{\rm QCD}$. The homogeneous trajectory then obeys
\begin{equation}
 \ddot{\overline X}+3H\dot{\overline X}
 +\lambda_\Phi(\overline X^2-f_R^2)\overline X=0.
 \label{eq:model_background}
\end{equation}
The second term describes Hubble damping, while the last term drives the field towards the moving minimum. Radial oscillations develop when the field can no longer follow that minimum adiabatically. We evolve the signed variable $\overline X$, allowing the trajectory to cross the origin.

The transverse fluctuation $y_k$, with comoving wavenumber $k$, evolves on this background according to
\begin{equation}
 \ddot y_k+3H\dot y_k+
 \left[\frac{k^2}{a^2}
 +\lambda_\Phi(\overline X^2-f_R^2)\right]y_k=0.
 \label{eq:model_modes}
\end{equation}
The positive momentum term $k^2/a^2$ prevents tachyonic growth at sufficiently large physical momenta. For sufficiently small $k/a$, the effective squared frequency can become negative during radial motion, producing tachyonic growth. Repeated oscillations of the background can also drive parametric resonance. At zero momentum, the exact solution $y_0=\delta\theta_0\overline X$, with constant $\delta\theta_0$, describes a uniform PQ rotation and provides a check on the numerical evolution.

The enlarged inflationary radius suppresses the angular perturbation associated with a fixed canonical fluctuation. We adopt the Planck reference scale $k_*=0.05\,\mathrm{Mpc}^{-1}$~\cite{Planck:2018jri} and denote its horizon-exit time by $t_*$, with $H_*=H(t_*)$. For a light angular mode and a slowly tracking radial background, the primordial phase power is
\begin{equation}
 \mathcal P_{\delta\theta}(k_*)
 =\left[\frac{H_*}{2\pi f_R(t_*)}\right]^2.
 \label{eq:model_primordial_phase}
\end{equation}
The factor $f_R^{-2}$ converts the transverse field fluctuation into an angular fluctuation. Once the QCD potential becomes important, this angular perturbation changes the local misalignment abundance. Its contribution to the cold-dark-matter isocurvature power is
\begin{equation}
 \mathcal P_S(k_*)=r_a^2
 \left(\frac{\partial\ln\Omega_a}{\partial\theta_i}\right)^2
 \mathcal P_{\delta\theta}(k_*).
 \label{eq:model_CMB_power}
\end{equation}
Here $\theta_i$ is the homogeneous misalignment angle and $r_a=\Omega_a/\Omega_{\rm dm}$ is the axion fraction of dark matter. The quantities $\Omega_a$ and $\Omega_{\rm dm}$ are present densities in units of the critical density. The abundance derivative measures the response to a change in angle, while $r_a$ weights that response by the axion contribution to the total dark matter~\cite{Allali:2025pja}.

We use the Planck constraint on uncorrelated, nearly scale-invariant isocurvature~\cite{Planck:2018jri},
\begin{equation}
 \beta_{\rm iso}\equiv
 \frac{\mathcal P_S}{A_s+\mathcal P_S}<0.038,
 \qquad A_s=2.1\times10^{-9},
 \label{eq:model_Planck}
\end{equation}
Here $\beta_{\rm iso}$ is the isocurvature fraction, and $A_s=2.1\times10^{-9}$ is the Planck normalization of the primordial curvature power spectrum at $k_*=0.05\,\mathrm{Mpc}^{-1}$~\cite{Planck:2018vyg}. For the benchmarks, axions account for all dark matter, $r_a=1$. In the harmonic approximation, we obtain the initial misalignment angle from the radiation-era abundance relation of Ref.~\cite{Graham:2025iwx},
\begin{equation}
 \theta_i=
 \left(\frac{7.4\times10^{11}\,\mathrm{GeV}}{f_a}\right)^{7/12}
 =0.0572
 \quad\text{at }f_a=10^{14}\,\mathrm{GeV}.
 \label{eq:benchmark_misalignment}
\end{equation}
In this regime $\Omega_a\propto\theta_i^2$, so the logarithmic derivative in Eq.~\eqref{eq:model_CMB_power} is $2/\theta_i$~\cite{Wantz:2009it,PhysRevD.33.889}. Each sign reversal of the homogeneous field changes its phase by $\pi$ relative to the fixed QCD minimum. We therefore choose the initial orientation for each trajectory to reach the stated final angle. The small correction to this abundance normalization from decay radiation is assessed below.

To follow the growth of finite-wavelength fluctuations, we define the transverse power and its integrated variance over the resolved momentum band,
\begin{equation}
 \Delta_Y^2=\frac{k^3|y_k|^2}{2\pi^2},
 \qquad
 \mathcal V_{\rm band}=
 \frac{1}{\pi^2f_a^2}
 \int_{k_{\min}}^{k_{\max}}d\ln k\,\Delta_Y^2.
 \label{eq:model_variance}
\end{equation}
Here $k_{\min}$ and $k_{\max}$ are the smallest and largest comoving wavenumbers included in the calculation. The quantity $\Delta_Y^2$ is the field variance per logarithmic momentum interval, and $\mathcal V_{\rm band}$ measures the integrated variance in units of $\pi^2f_a^2$. We stop an integration when $\mathcal V_{\rm band}$ first reaches $10^{-2}$, corresponding to a root-mean-square transverse fluctuation of $0.314f_a$. Throughout the calculation, the PQ sector is treated as a spectator and its potential is assumed to remain quartic over the field excursion. In Appendix~\ref{app:background_methods}, we verify that the PQ field contributes only small corrections to the background expansion and the effective Planck mass. Appendix~\ref{app:numerical_validation} compares the fluctuation contributions to the radial force and energy with those of the homogeneous condensate.

\section{Saxion decay and thermal effects}
\label{sec:decay}

The KSVZ completion introduces one Dirac colour triplet $\psi$, whose mass originates from PQ symmetry breaking~\cite{DiLuzio:2020wdo}. With real Yukawa coupling $y_K$, its Lagrangian is
\begin{equation}
 \mathcal L_{\rm KSVZ}=
 \overline\psi i\slashed D\psi
 -\left(y_K\Phi\overline\psi_L\psi_R+\mathrm{h.c.}\right)
 +\mathcal L_{\rm mix},
 \label{eq:ksvz_lagrangian}
\end{equation}
where $D_\mu$ is the gauge-covariant derivative and $\psi_L$ and $\psi_R$ are the left- and right-handed components of $\psi$. The interaction $\mathcal L_{\rm mix}$ preserves PQ symmetry and permits the heavy quark to decay into Standard Model particles; its form is specified in Appendix~\ref{app:micro_decay}.

We write the PQ radius as $\rho=f_a+s$, where the saxion $s$ describes radial fluctuations about the vacuum. The Yukawa interaction generates the fermion mass $m_\psi=y_Kf_a/\sqrt2$ and couples the saxion to the fermion with strength $g_Y=y_K/\sqrt2$. For colour multiplicity $N_c=3$, the vacuum width into a fermion pair is~\cite{Benabou:2023ghl}
\begin{equation}
 \Gamma^{(0)}_{\psi\bar\psi}
 =\frac{3y_K^2m_s}{16\pi}(1-u)^{3/2}\Theta(1-u),
 \qquad
 u\equiv\frac{y_K^2}{\lambda_\Phi}
 =\frac{4m_\psi^2}{m_s^2}.
 \label{eq:ksvz_vacuum_width}
\end{equation}
The superscript $(0)$ denotes decay widths evaluated in vacuum. The step function enforces the threshold $m_s>2m_\psi$, while $(1-u)^{3/2}$ describes the suppression as the threshold is approached. Increasing $y_K$ strengthens the vertex but also increases the daughter masses, so the width has a maximum within the open-channel region.

The saxion also couples to axions through the radial dependence of the angular kinetic term. For negligible axion mass, this channel has width
\begin{equation}
 \Gamma^{(0)}_{aa}
 =\frac{m_s^3}{32\pi f_a^2}
 =\frac{\lambda_\Phi m_s}{16\pi}.
 \label{eq:radial_axion_width}
\end{equation}
Thus the axion width is fixed by the quartic coupling and the vacuum saxion mass. Within the kinematically allowed range $0<u<1$, the fermion width is bounded by
\begin{equation}
 \frac{\Gamma^{(0)}_{\psi\bar\psi}}{m_s}
 \lesssim 1.11\times10^{-2}\lambda_\Phi.
 \label{eq:ksvz_width_bound}
\end{equation}
Both widths are therefore small compared with the oscillation frequency when $\lambda_\Phi\ll1$.

The dimensionless ratio $u=y_K^2/\lambda_\Phi=4m_\psi^2/m_s^2$ determines how close the fermion channel lies to its decay threshold, with the channel open for $0<u<1$. At fixed $\lambda_\Phi$ and $f_a$, increasing $u$ strengthens the Yukawa interaction but also raises the fermion mass, reducing the available phase space. The resulting dependence $\Gamma_{\psi\bar\psi}^{(0)}\propto u(1-u)^{3/2}$ reaches its maximum at $u=2/5$.

Within the two-channel approximation, we define $\Gamma_s=\Gamma^{(0)}_{aa}+\Gamma^{(0)}_{\psi\bar\psi}$ and $B_i=\Gamma_i^{(0)}/\Gamma_s$. Since the axion width is independent of $u$ at fixed $\lambda_\Phi$ and $f_a$, the same choice maximizes the fermion branching fraction and minimizes the axion branching fraction,
\begin{equation}
 B_{\psi\bar\psi}\lesssim0.358,
 \qquad
 B_{aa}\gtrsim0.642.
 \label{eq:two_channel_branching}
\end{equation}
We therefore adopt $u=2/5$ for the reference histories to maximize the fraction of decay energy transferred to the visible sector. Even with this choice, most of the saxion energy goes into axions.

We neglect Higgs-portal decays in the two-channel calculation. This approximation applies when the renormalized portal coupling satisfies $|\lambda_{\Phi H}|\ll0.1\lambda_\Phi$, so that $\Gamma_{\mathcal H\mathcal H^\dagger}/\Gamma_{aa}^{(0)} =(\lambda_{\Phi H}/\lambda_\Phi)^2\ll0.01$. Here $\mathcal H$ denotes the Standard Model Higgs doublet. Additional decay channels depend on the particle-physics completion~\cite{Fan:2011ua}, with gauge-loop channels suppressed by a factor of order $\alpha_s^2/\pi^2$ relative to the axion channel. Appendix~\ref{sec:portal_assumption} checks that the portal thermal force and the portal coupling induced by quark mixing remain small under these assumptions.

The Yukawa interaction also produces a thermal force on the radial field. For KSVZ quarks in equilibrium with $m_\psi(\rho)=y_K\rho/\sqrt2\ll T$, the leading field-dependent free energy is $\Pi_\psi\rho^2/2$, with~\cite{Mukaida:2012qn,Quiros:1999jp}
\begin{equation}
 \Pi_\psi=\frac{N_cy_K^2}{12}T^2
 \equiv\alpha_\psi T^2.
 \label{eq:positive_thermal_curvature}
\end{equation}
Here $T$ is the bath temperature, $N_c=3$ counts colours and $\alpha_\psi$ is the dimensionless thermal-mass coefficient. This positive term pulls the radial minimum towards the origin, opposing its curvature-induced displacement. As $m_\psi(\rho)/T$ increases, the declining quark population weakens this thermal force. Appendix~\ref{app:micro_thermal} retains the full mass dependence and evaluates the renormalized fermion vacuum correction~\cite{PhysRevD.7.1888,Quiros:1999jp}.

For the reference histories, the small thermal and vacuum corrections support the tree-level force used in Eqs.~\eqref{eq:model_background} and \eqref{eq:model_modes}. Energy transfer through cold nonadiabatic fermion production is estimated separately in Appendix~\ref{app:micro_decay}. Appendix~\ref{app:micro_response} tests collisional damping and the associated thermal fluctuations using the local response of Refs.~\cite{Mukaida:2012qn,Moroi:2013tea,Moroi:2014mqa}. The vacuum decay widths apply at the later decay epoch, when the radial motion is approximately harmonic and the bath has cooled below the saxion and fermion masses.

\section{PQ field evolution after inflation}
\label{sec:numerics}

We now follow the PQ condensate through the transition from inflation to radiation domination. To examine how the transition duration affects its evolution, we hold the inflationary normalization fixed and vary the interval over which the curvature decreases. We describe expansion using $N=\ln(a/a_e)$, where $a_e$ is the scale factor at the centre of the transition, corresponding to $N=0$. A prime denotes differentiation with respect to $N$.

A smooth interpolation between the early and late stages is specified by
\begin{equation}
 L(N)=\frac{1}{1+e^{-N/\Delta}},
 \qquad \Delta>0.
 \label{eq:num_release_function}
\end{equation}
The function rises from zero to unity, and $\Delta$ controls its width in e-folds. To retain slow-roll evolution well before the transition, we introduce
\begin{equation}
 \epsilon_{\rm sr}(N)=\epsilon_0L^{b\Delta}.
 \label{eq:num_slowroll_function}
\end{equation}
At early times, this reduces to $\epsilon_{\rm sr}\simeq\epsilon_0e^{bN}$. The parameter $b$ controls its evolution and hence the early scalar tilt, while $\epsilon_0$ is fixed by the scalar power normalization.

We prescribe the logarithmic change of the Hubble rate as
\begin{equation}
 \epsilon_H\equiv-\frac{H'}H
 =2L+\epsilon_{\rm sr}(1-L).
 \label{eq:num_reconstructed_functions}
\end{equation}
This approaches the slow-roll form during inflation and the radiation value $\epsilon_H=2$ at late times. The radiation density $\rho_r$ is specified through its fraction of the background energy,
\begin{equation}
 r_r\equiv\frac{\rho_r}{3M_{\rm P}^2H^2}=L^2.
 \label{eq:num_radiation_fraction}
\end{equation}
Radiation therefore becomes dominant as $L\to1$. At the same time, the curvature $R=6(2-\epsilon_H)H^2$ tends to zero and the PQ minimum contracts towards $f_a$. A larger $\Delta$ allows more time for the radial field to follow this motion.

In Appendix~\ref{app:background_methods}, we reconstruct a canonical, minimally coupled inflaton that reproduces this expansion history. Its kinetic and potential energies remain positive during the transition, while energy flows from the inflaton into radiation. This construction defines the smooth reheating family used in the calculation. When both the inflaton and the PQ field couple nonminimally to gravity, their couplings can modify the inflationary isocurvature constraint~\cite{Rigouzzo:2025ycb}. After inflation, oscillations in the spacetime curvature can periodically vary the effective PQ mass, driving additional radial motion and amplifying field fluctuations~\cite{Graham:2025iwx}.

We adopt $H_*=1.8\times10^9\,\mathrm{GeV}$ as a reference inflationary Hubble rate and keep it fixed when comparing different PQ couplings and curvature-release durations. For the PQ parameters specified below, this choice places the inflationary radius well above its vacuum value. We choose $b=0.035$ to reproduce the scalar spectral index $n_s\simeq0.965$ of the Planck 2018 baseline analysis~\cite{Planck:2018vyg}. Let $N_*>0$ denote the interval between horizon exit of $k_*$ and $N=0$. During early inflation, $d\ln\epsilon_H/dN\simeq b$, so the first-order relation $n_s-1\simeq-2\epsilon_{H,*}-b$ reduces to $n_s\simeq1-b$ for the small horizon-exit value $\epsilon_{H,*}\equiv\epsilon_H(-N_*)$ used here.

The observed scalar amplitude $A_s=2.1\times10^{-9}$ determines $\epsilon_{H,*}=H_*^2/(8\pi^2M_{\rm P}^2A_s)\simeq3.30\times10^{-12}$. Combining this normalization with matching the scale $k_*$ through the subsequent entropy history determines both the interval $N_*$ from horizon exit to $N=0$ and the coefficient $\epsilon_0$. For the chosen release width $\Delta=0.20$, we find $N_*=50.85$ and $H_e\equiv H(0)=1.364\times10^9\,\mathrm{GeV}$, with $N=0$ marking the centre of the transition.

The effective number of relativistic energy degrees of freedom is $g_*=106.75+(7/8)\times12=117.25$, where the Standard Model contributes $106.75$ and the Dirac KSVZ quark has 12 degrees of freedom from spin, colour and antiparticles. Using this count in the prescribed radiation density, we obtain a maximum bath temperature $T_{\max}=2.30\times10^{13}\,\mathrm{GeV}$. The same inflationary normalization predicts the tensor-to-scalar ratio $r_{\rm tens}=16\epsilon_{H,*}=5.27\times10^{-11}$. The temperature, curvature and tensor-to-scalar ratio therefore follow from the chosen background and its scalar normalization.

For the PQ benchmarks, we take $f_a=10^{14}\,\mathrm{GeV}$, $\xi=1$ and $u=2/5$. We start the homogeneous field at the slowly moving inflationary minimum, with $\overline X=f_R$ and $\dot{\overline X}=\dot f_R$. We initialize the transverse fluctuations in the adiabatic vacuum at wavelengths much shorter than the Hubble radius. Writing their comoving momenta as $\kappa=k/(a_eH_e)$, we evolve modes over $0.1\leq\kappa\leq10$. We describe the mode normalization, convergence tests and evolution of longitudinal fluctuations in Appendix~\ref{app:numerical_validation}.

Figure~\ref{fig:pq_relaxation} compares the radial trajectories and transverse amplification for the benchmarks A--C. At smaller $\lambda_\Phi$, the inflationary radius is larger and the field undergoes a longer sequence of origin crossings. A and B eventually settle into a broken minimum, whereas C reaches the fluctuation cutoff at $N\simeq3.30$. Thus the stronger suppression of primordial isocurvature in C accompanies greater amplification during relaxation.

\begin{figure}[tbp]
 \centering
 \includegraphics[width=\textwidth]{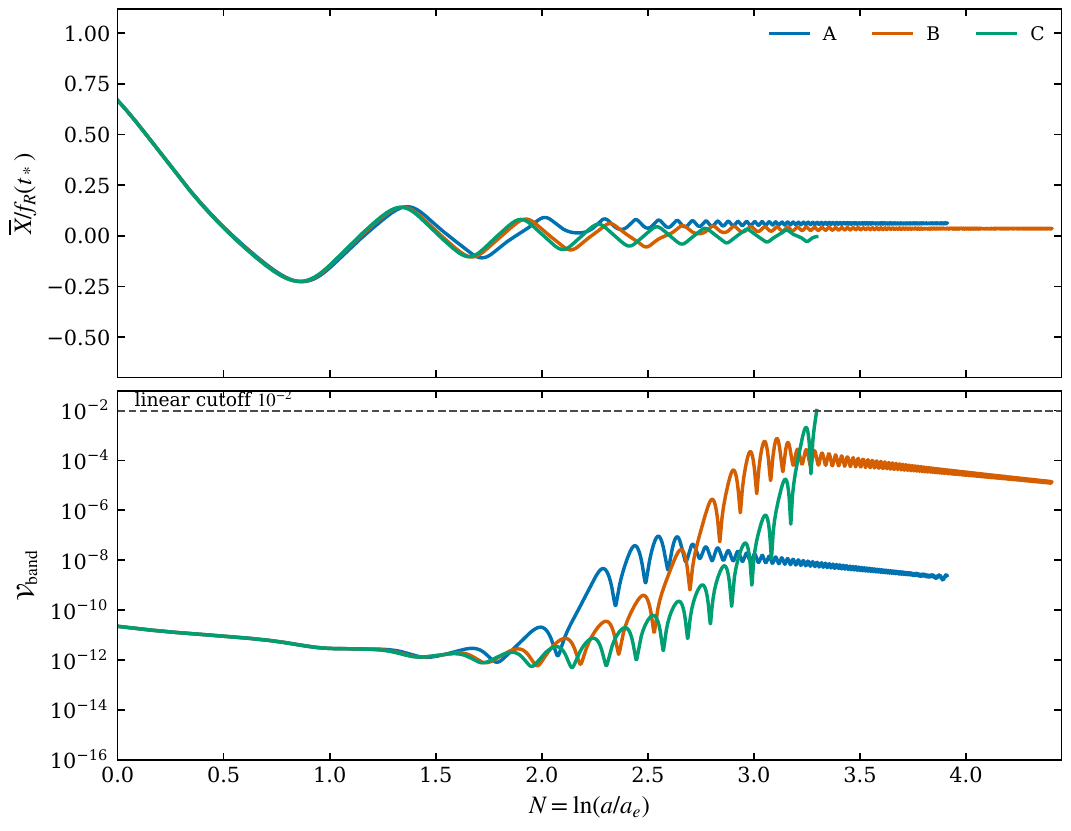}
\caption{Curvature-driven relaxation of benchmarks A (blue), B (orange) and C (green) at $\Delta=0.20$, with the remaining parameters in Table~\ref{tab:cosmological_benchmarks}. The upper panel shows the signed homogeneous field $\overline X/f_R(t_*)$ against $N=\ln(a/a_e)$. As the curvature decreases, the radial minimum moves towards the vacuum. The field overshoots this moving minimum and passes through the origin, while expansion gradually damps its oscillations. The lower panel shows the transverse variance $\mathcal V_{\rm band}$ on a logarithmic scale, integrated over $0.1\leq\kappa\leq10$ with $\kappa=k/(a_eH_e)$. Its rapid growth reflects tachyonic and resonant amplification driven by the radial motion. The smaller quartic couplings in B and C prolong the sequence of origin crossings and produce stronger amplification. Once A and B settle near a broken minimum, amplification weakens and expansion dilutes the fluctuations. Both approach the same vacuum radius $|\overline X|=f_a$; their different late offsets arise from normalization by their respective inflationary radii $f_R(t_*)$. We stop the linear evolution of benchmark C when its variance first reaches $\mathcal V_{\rm band}=10^{-2}$, marked by the filled point on the dashed line.}
\label{fig:pq_relaxation}
\end{figure}

Table~\ref{tab:cosmological_benchmarks} summarizes the reference parameters and the late abundances calculated in Section~\ref{sec:results}. Here $\rho_s$ is the coherent saxion energy density, $\rho_{\rm tot}$ is the total density, and $\Delta N_{\rm eff}$ measures the additional radiation in units of one neutrino species. The angular spectra in Fig.~\ref{fig:angular_spectra} show the amplified momentum range and the wave energy remaining after A and B settle.

\begin{table}[tbp]
 \centering\small
 \setlength{\tabcolsep}{4pt}
 \begin{tabular}{lrrrrrr}
 \toprule
 Benchmark point & $\lambda_\Phi$ & $f_R(t_*)/f_a$ & $\beta_{\rm iso}$ & $\max\mathcal V_{\rm band}$ & \shortstack{$\max$\\$\rho_s/\rho_{\rm tot}$} & $\Delta N_{\rm eff}$\\
 \midrule
 A & $3\times10^{-11}$ & $16.1$ & $0.0181$ & $9.2\times10^{-8}$ & $0.0087$ & $0.033$\\
 B & $10^{-11}$ & $27.9$ & $0.00611$ & $7.6\times10^{-4}$ & $0.031$ & $0.12$\\
 C & $3\times10^{-12}$ & $50.9$ & $0.00184$ & $10^{-2}$ (stop) & --- & ---\\
 \bottomrule
 \end{tabular}
 \caption{Reference histories at $\Delta=0.20$, $f_a=10^{14}\,\mathrm{GeV}$, $\xi=1$ and $u=2/5$. The isocurvature values use Eq.~\eqref{eq:benchmark_misalignment}, and the variance covers $0.1\leq\kappa\leq10$. Late abundances follow from action matching and two-channel decay. The evolution of C ends at the fluctuation cutoff.}
 \label{tab:cosmological_benchmarks}
\end{table}

The thermal correction to the PQ potential remains small in these histories, with $\Pi_\psi/(\lambda_\Phi f_a^2)\leq5.3\times10^{-3}$. When we include this term in both the homogeneous and transverse field equations, the peak variance of benchmark B changes by only $0.024\%$. The PQ field also has little effect on the background expansion. The amplified transverse PQ field fluctuations can nevertheless alter the radial motion while carrying only a small fraction of the total cosmological energy. We therefore assess their backreaction relative to the homogeneous condensate in Appendix~\ref{app:numerical_validation} and discuss the resulting uncertainty in the surviving saxion abundance in Section~\ref{sec:discussion}.

\begin{figure}[tbp]
 \centering
 \includegraphics[width=\textwidth]{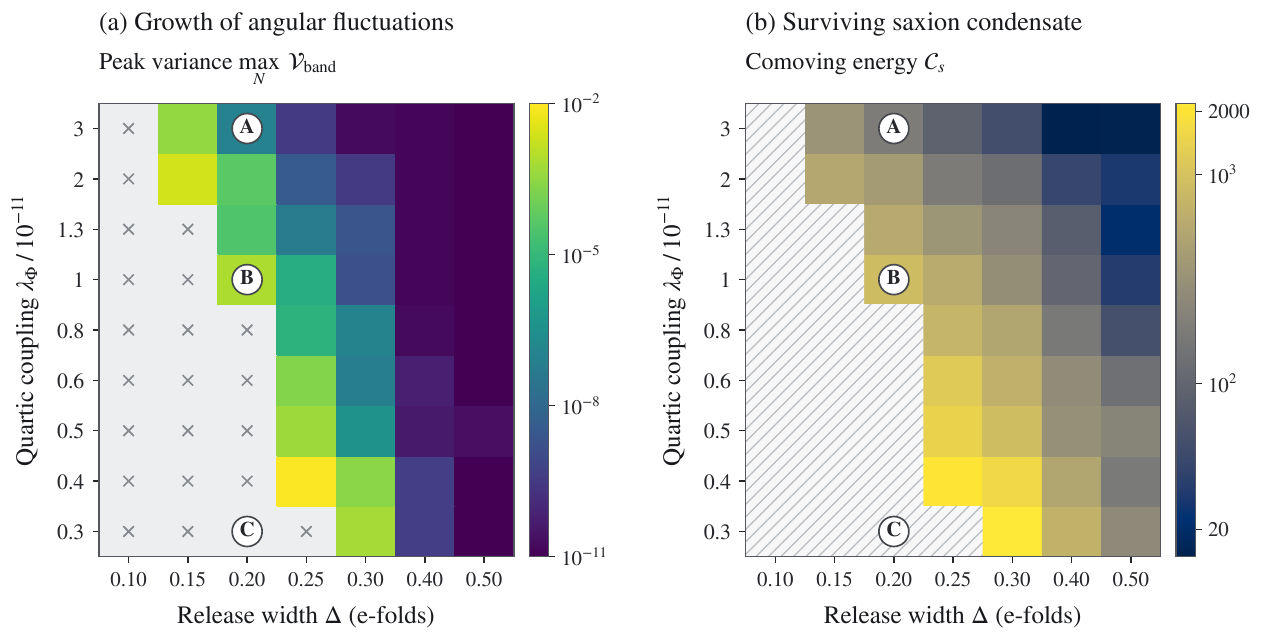}
\caption{PQ relaxation for 63 pairs of release width $\Delta$ and quartic coupling $\lambda_\Phi$, at fixed $H_*$, $f_a$, $\xi$ and $u$. Both panels use the same parameter axes, so corresponding cells represent the same history. The left panel shows the maximum transverse variance $\mathcal V_{\rm band}$ over $0.1\leq\kappa\leq10$, where $\kappa=k/(a_eH_e)$, measuring the angular fluctuations excited during radial relaxation. The right panel shows the surviving comoving saxion energy $\mathcal C_s=e^{3N_m}\rho_{s,m}/(\lambda_\Phi f_a^4)$, evaluated using the matching condition in Eq.~\eqref{eq:robustness_matching}. Each panel has a separate logarithmic colour scale because fluctuation variance and condensate energy are different quantities with different numerical ranges. Brighter colours indicate stronger fluctuations on the left and more surviving normalized energy on the right. Gradual curvature release allows the field to follow its moving minimum more closely, reducing fluctuation growth and, at fixed $\lambda_\Phi$, the coherent energy available for late saxion decay. Gray cells with crosses reach the cutoff $\mathcal V_{\rm band}=10^{-2}$; the corresponding hatched cells have undetermined late abundances because evolution ends before matching. Circled letters identify benchmarks A, B and C. The smallest variances approach the vacuum contribution within the resolved momentum interval.}
 \label{fig:release_robustness}
\end{figure}

To examine the dependence on release duration, Fig.~\ref{fig:release_robustness} compares 63 pairs $(\lambda_\Phi,\Delta)$ at the same inflationary normalization. Of these, 22 reach the transverse cutoff, while 41 reach the matching epoch with the variance remaining below it. Every sampled quartic reaches the cutoff at the fastest release, $\Delta=0.10$, whereas all sampled histories with $\Delta\geq0.30$ remain below it through the matching epoch. The strong suppression of amplification at longer release duration reflects the improved tracking of the contracting minimum.

In the right panel of Fig.~\ref{fig:release_robustness}, we show the coherent energy remaining after relaxation. Once the radial oscillations are rapid compared with expansion and approximately harmonic, their cycle-averaged energy density redshifts as $\rho_s\propto a^{-3}$, provided energy loss through saxion decay remains negligible~\cite{Masso:2005zg}. We remove this leading dilution by defining the dimensionless comoving saxion energy at the matching time $N_m$ as
\begin{equation}
 \mathcal C_s
 = \frac{e^{3N_m}\rho_{s,m}}{\lambda_\Phi f_a^4},
\end{equation}
where $\rho_{s,m}$ is the homogeneous kinetic plus vacuum potential energy density at matching. Since $N=\ln(a/a_e)$, the factor $e^{3N_m}$ compensates for the matter-like dilution, while $\lambda_\Phi f_a^4$ supplies the energy-density normalization.

For each pair $(\lambda_\Phi,\Delta)$ that remains below the fluctuation cutoff, we evaluate $\mathcal C_s$ at the common small-amplitude condition in Eq.~\eqref{eq:robustness_matching}. To determine the initial saxion abundance for the late decay calculation, we average the comoving oscillation action over complete cycles, reducing the dependence on the oscillation phase at matching. We describe both matching prescriptions and test their stability in Appendix~\ref{app:numerical_validation}.

\section{Saxion decay and dark radiation}
\label{sec:results}

When the field settles into small oscillations about a broken minimum, it moves in an approximately quadratic potential, $m_s^2s^2/2$. At this stage, $m_s\gg H,\Gamma_s$, so expansion and saxion decay change the amplitude very little during a single oscillation. The kinetic and potential energies are then nearly equal when averaged over a cycle, leaving negligible average pressure. We can therefore treat the surviving condensate as a pressureless fluid and follow its evolution without resolving each oscillation~\cite{PhysRevD.28.1243,Masso:2005zg}. To obtain the initial fluid density from the field trajectory, we use the oscillation action per unit physical volume,
\begin{equation}
 J=\frac{1}{2\pi}\oint\dot{\overline X}\,d\overline X,
 \qquad
 \rho_{s,m}=m_sJ_m,
 \label{eq:main_action_matching}
\end{equation}
where the integral follows one complete oscillation and the subscript $m$ denotes the matching time between the field and fluid descriptions. When expansion is slow over a period and saxion decay removes negligible energy, the comoving action $a^3J$ is approximately conserved~\cite{Masso:2005zg}. In the harmonic limit, $J=\rho_s/m_s$ can be interpreted as an effective saxion number density. Conservation of $a^3J$ then corresponds to a fixed effective number of saxions in a comoving volume, with their density decreasing as $a^{-3}$. The relation $\rho_{s,m}=m_sJ_m$ thus connects the surviving radial oscillations directly to the initial conditions for the late decay calculation.

We average $a^3J$ over six complete cycles to reduce the dependence on oscillation phase. At the reference matching times, the oscillation amplitudes are approximately $0.04f_a$, while $m_s/H$ is of order $10^3$. At these amplitudes the motion is approximately harmonic, and the oscillation period is short compared with the expansion time. The cycle-averaged pressure is therefore small compared with the saxion energy density. Appendix~\ref{app:numerical_validation} describes the quartic action and the matching tests.

The averaged saxion density obeys~\cite{Bae:2014rfa}
\begin{equation}
 \dot\rho_s+3H\rho_s=-\Gamma_s\rho_s.
 \label{eq:late_saxion_fluid}
\end{equation}
The expansion term describes matter-like dilution, while saxion decay transfers energy out of the condensate at a rate $\Gamma_s\rho_s$ per unit physical volume. The energy density of relativistic axions produced by saxion decay evolves according to
\begin{equation}
 \dot\rho_{a,\rm rel}+4H\rho_{a,\rm rel}
 =\Gamma^{(0)}_{aa}\rho_s.
 \label{eq:late_receiving_fluids}
\end{equation}
Here $\rho_{a,\rm rel}$ denotes the component produced by saxion decay. Its energy redshifts as radiation after injection.

The fermion channel transfers energy to the visible bath, whose density and pressure are denoted by $\rho_v$ and $p_v$. Its continuity equation is
\begin{equation}
 \dot\rho_v+3H(\rho_v+p_v)
 =\Gamma^{(0)}_{\psi\bar\psi}\rho_s.
 \label{eq:late_visible_fluid}
\end{equation}
The KSVZ mixing interaction allows the decay products to deposit their energy rapidly into the visible sector. We retain the temperature-dependent equation of state of the equilibrium KSVZ quark, together with 106.75 relativistic Standard Model degrees of freedom. This accounts for the disappearance of thermal quarks as the bath cools below their mass.

The expansion rate follows from the combined energy density,
\begin{equation}
 3M_{\rm P}^2H^2
 =\rho_s+\rho_{a,\rm rel}+\rho_v
 \equiv\rho_{\rm tot}.
 \label{eq:late_Friedmann}
\end{equation}
Since $\Gamma_s=\Gamma^{(0)}_{aa}+\Gamma^{(0)}_{\psi\bar\psi}$, the decay sources cancel when the three continuity equations are added. The energy lost through saxion decay is therefore accounted for in the axion and visible components as the Universe expands.

The total width also determines when this energy transfer becomes important. We characterize that epoch by $T_\Gamma\equiv T(H=\Gamma_s)$, when the saxion lifetime becomes comparable to the expansion time~\cite{Jeong:2013oza}. For the benchmarks A and B, visible radiation dominates the energy density, so we can estimate this temperature by substituting $\rho_v\simeq\pi^2g_*T^4/30$ into Eq.~\eqref{eq:late_Friedmann},
\begin{equation}
 H(T)\simeq\sqrt{\frac{\pi^2g_*}{90}}\frac{T^2}{M_{\rm P}},
 \qquad
 T_\Gamma\simeq
 \left(\frac{90}{\pi^2g_*}\right)^{1/4}
 \sqrt{\Gamma_sM_{\rm P}}.
 \label{eq:late_decay_temperature}
\end{equation}
Here we use $g_*=106.75$, since the equilibrium KSVZ quark population is negligible at these temperatures. Evaluating Eqs.~\eqref{eq:ksvz_vacuum_width} and \eqref{eq:radial_axion_width} for the benchmark parameters, we obtain total widths of $\Gamma_s\simeq7.20\times10^{-4}\,\mathrm{GeV}$ for A and $1.39\times10^{-4}\,\mathrm{GeV}$ for B. Equation~\eqref{eq:late_decay_temperature} then predicts $T_\Gamma\simeq2.26\times10^7\,\mathrm{GeV}$ and $9.93\times10^6\,\mathrm{GeV}$ for A and B, respectively.\footnote{The full fluid evolution gives $T_\Gamma\simeq2.26\times10^7\,\mathrm{GeV}$ for A and
$9.83\times10^6\,\mathrm{GeV}$ for B, in close agreement with these estimates.} In both benchmarks, $T_\Gamma/m_s<0.03$ and $T_\Gamma/m_\psi<0.10$, so the bath is substantially colder than the saxion and KSVZ quark during the main decay epoch. This supports the use of vacuum decay widths. We describe the initial energy matching and temperature evolution in Appendix~\ref{app:numerical_validation}.

Figure~\ref{fig:late_energy_transfer} shows the resulting energy fractions. Before the decay becomes important, the saxion fraction increases because matter dilutes more slowly than radiation. Then the saxion decay removes the condensate and leaves a freely propagating axion population. Benchmark B begins with more surviving coherent energy than A and reaches a larger saxion fraction, although radiation dominates the expansion throughout both histories.

\begin{figure}[tbp]
 \centering
 \includegraphics[width=\textwidth]{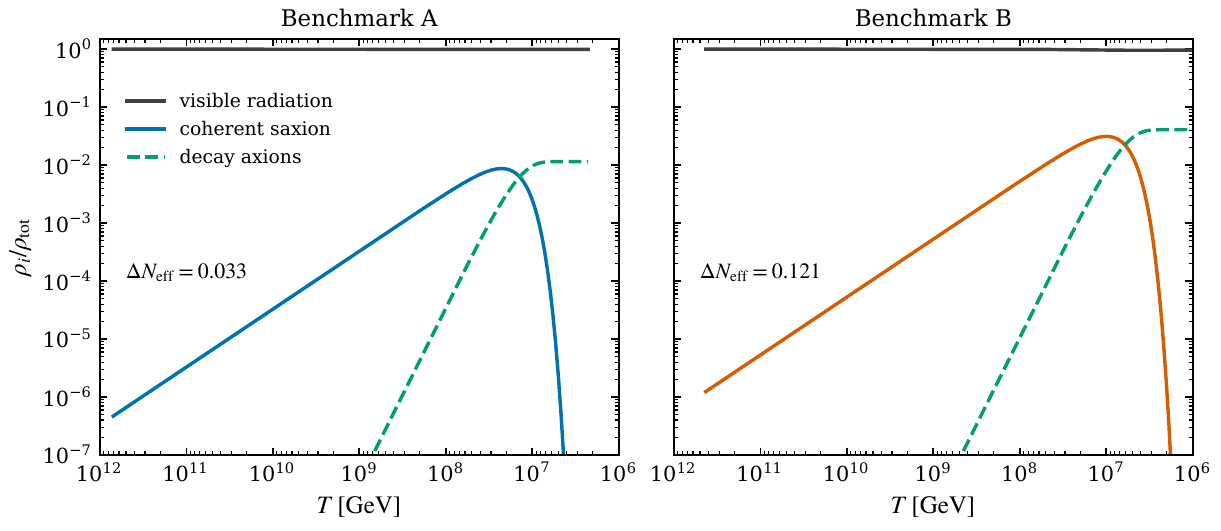}
 \caption{Energy fractions $\rho_i/\rho_{\rm tot}$ during late saxion decay for benchmark A (left) and benchmark B (right), with $u=2/5$, $B_{aa}=0.642$ and $B_{\psi\bar\psi}=0.358$. Blue and orange curves represent the coherent saxion and decay-produced relativistic axions, respectively; the dashed gray curve represents the visible bath. Both axes are logarithmic, with temperature in GeV decreasing from left to right. The visible equation of state includes the equilibrium massive KSVZ quark and 106.75 relativistic Standard Model degrees of freedom. The nearly horizontal visible fraction reflects the large pre-existing bath and the normalization by the evolving total density, even as decay increases its comoving entropy. The axion fraction approaches a plateau once energy transfer ends and both radiation components redshift approximately as $a^{-4}$.}
 \label{fig:late_energy_transfer}
\end{figure}

After energy transfer is complete, the axion-to-visible energy ratio determines the additional effective neutrino number,
\begin{equation}
 \Delta N_{\rm eff}=\frac{43}{7}
 \left(\frac{10.75}{g_{*s,d}}\right)^{1/3}
 \left.\frac{\rho_{a,\rm rel}}{\rho_v}\right|_d,
 \qquad
 g_{*s,d}=106.75.
 \label{eq:late_Neff}
\end{equation}
The epoch $d$ is chosen after decay and the disappearance of the KSVZ quarks, and $g_{*s}$ counts visible entropy degrees of freedom~\cite{Moroi:2014mqa}. The factor $43/7$ expresses the radiation ratio at neutrino decoupling in units of one neutrino species. The cube-root factor accounts for the heating of the visible bath relative to the decoupled axions as massive particles disappear
between saxion decay and neutrino decoupling.

The resulting abundances are $\Delta N_{\rm eff}\simeq0.033$ for A and $0.12$ for B, as listed in Table~\ref{tab:cosmological_benchmarks}. These values describe radiation from coherent-condensate decay; thermal axion and saxion populations are estimated separately in Appendix~\ref{sec:benchmark_thermal_estimates}. The decay axions remain relativistic to the present, and their scattering rate at injection is small compared with $H$. The large radial mass during inflation suppresses independent fluctuations of the condensate on CMB scales. Under the assumed single-clock reheating history, the decay radiation is therefore adiabatic on these scales. We estimate the present momentum of the decay axions in Appendix~\ref{app:numerical_validation} and discuss their scattering on the thermal bath, together with the thermal production of axions and saxions, in Appendix~\ref{app:micro_response}.

For a subdominant saxion condensate, we can estimate the final axion radiation density analytically. Let $r_m=\rho_{s,m}/\rho_{v,m}$ and $H_m=H(N_m)$, with $H_m\gg\Gamma_s$. For radiation domination and constant relativistic degrees of freedom, integration of the decay source gives, to leading order in the condensate fraction,
\begin{equation}
 \left.\frac{\rho_{a,\rm rel}}{\rho_v}\right|_d
 \simeq B_{aa}r_m
 \sqrt{\frac{\pi H_m}{2\Gamma_s}}.
 \label{eq:analytic_rd_abundance}
\end{equation}
The factor $r_m$ specifies the initial saxion energy relative to visible radiation. The square-root factor accounts for the growth of this ratio before decay. A longer lifetime allows the matter-like condensate to become more important compared with the radiation bath~\cite{Jeong:2013oza}.

The visible decay channel affects the axion radiation abundance through both the branching fraction and the saxion lifetime. For an open fermion channel, $0<u<1$, taking the ratio of the vacuum widths in Eqs.~\eqref{eq:ksvz_vacuum_width} and
\eqref{eq:radial_axion_width}~\cite{Mazumdar:2016nzr,Benabou:2023ghl},
we obtain
$R_\Gamma=\Gamma^{(0)}_{\psi\bar\psi}/\Gamma^{(0)}_{aa}
=3u(1-u)^{3/2}$.
The axion branching fraction and total width are then
$B_{aa}=(1+R_\Gamma)^{-1}$ and
$\Gamma_s=(1+R_\Gamma)\Gamma^{(0)}_{aa}$.
At fixed initial condensate abundance and axion width, Eq.~\eqref{eq:analytic_rd_abundance} therefore implies
\begin{equation}
 \Delta N_{\rm eff}\propto B_{aa}\Gamma_s^{-1/2}
 \propto(1+R_\Gamma)^{-3/2}.
 \label{eq:branching_lifetime_scaling}
\end{equation}
The factor $(1+R_\Gamma)^{-1}$ reflects the smaller fraction of decay energy transferred to axions. The additional factor $(1+R_\Gamma)^{-1/2}$ arises from the shorter saxion lifetime, which leaves less time for the condensate density to grow relative to radiation before it decays. Since $R_\Gamma$ is largest at $u=2/5$, this choice minimizes $\Delta N_{\rm eff}$ within the subdominant-condensate approximation. Setting $R_\Gamma=0$ increases the estimated abundance by a factor of approximately $1.94$ relative to $u=2/5$, for the same initial condensate and axion width.

In Table~\ref{tab:release_dark_radiation}, we show how the duration of curvature release affects the surviving saxion abundance and
the resulting dark radiation at fixed $\lambda_\Phi=10^{-11}$. We characterize the surviving condensate by the dimensionless comoving action $\mathcal C_J=e^{3N_m}m_sJ_m/(\lambda_\Phi f_a^4)$. As $\Delta$ increases from $0.20$ to $0.50$, the surviving action and the predicted radiation abundance decrease by factors close to $28$. The primordial phase power remains fixed, as do the particle masses and decay widths. The change in $\Delta N_{\rm eff}$ therefore measures the reduction in radial excitation left by the more gradual contraction of the minimum.

\begin{table}[tbp]
 \centering
 \small
 \begin{tabular}{crrrr}
 \toprule
 $\Delta$ & $\max\mathcal V_{\rm band}$ & $\mathcal C_J$ & $\max(\rho_s/\rho_{\rm tot})$ & $\Delta N_{\rm eff}$\\
 \midrule
 $0.20$ & $7.6\times10^{-4}$ & $876$ & $0.031$ & $0.12$\\
 $0.30$ & $1.6\times10^{-9}$ & $296$ & $0.011$ & $0.041$\\
 $0.40$ & $1.3\times10^{-11}$ & $100$ & $0.0037$ & $0.014$\\
 $0.50$ & $1.0\times10^{-11}$ & $31.6$ & $0.0012$ & $0.0044$\\
 \bottomrule
 \end{tabular}
 \caption{Dependence of the surviving action and late radiation abundance on release width at fixed $\lambda_\Phi=10^{-11}$ and the remaining benchmark parameters. Variance maxima cover $0\leq N\leq4.5$ and $0.1\leq\kappa\leq10$; the last two are set by the vacuum contribution. Each late abundance follows from the comoving action $\mathcal C_J$ and the same two-channel decay model.}
 \label{tab:release_dark_radiation}
\end{table}

The predicted $\Delta N_{\rm eff}$ depends only weakly on the matching epoch, with an uncertainty much smaller than the possible corrections from fluctuation backreaction discussed in Section~\ref{sec:discussion}. Residual saxion oscillations can continue to amplify axions through narrow parametric resonance as expansion redshifts modes through the resonance band. After subtracting the perturbative decay contribution already included in the fluid equations, we estimate that this additional production depletes the condensate by less than roughly one part in $10^4$ in A and B.

The fermion decay channel increases the visible comoving entropy by less than $2\%$ before QCD axion oscillations begin. The additional axion radiation also raises the expansion rate at a fixed bath temperature, delaying the onset of these oscillations. At a fixed initial angle, this increases the harmonic misalignment abundance by approximately $0.6\%$ in A and $2\%$ in B. We retain the common misalignment angle in Eq.~\eqref{eq:benchmark_misalignment} for the quoted isocurvature values and quantify these small corrections in Appendix~\ref{app:numerical_validation}.

For comparison with observations, adding the calculated axion radiation contribution to the Standard Model value $N_{\rm eff}^{\rm SM}=3.044$~\cite{Froustey:2020mcq} gives $N_{\rm eff}\simeq3.077$ for A and $3.165$ for B. The ACT DR6 analysis combined with Planck, CMB lensing and baryon acoustic oscillation data reports $N_{\rm eff}=2.86\pm0.13$ at 68\% confidence and a one-sided 95\% upper limit of $3.08$~\cite{AtacamaCosmologyTelescope:2025nti}. Benchmark A therefore lies just below the published bound, whereas the prediction for B exceeds it and is disfavoured under the assumptions of that analysis. A formal assessment of our full cosmological model would require a joint treatment of the physical prior $\Delta N_{\rm eff}\geq0$ and the accompanying cold-dark-matter isocurvature, together with a calculation of the nonlinear correction to B's condensate abundance. The smaller radiation contributions at $\Delta=0.40$ and $0.50$ in Table~\ref{tab:release_dark_radiation} fall below the quoted bound while retaining the same primordial angular normalization as B.

\section{Discussion}
\label{sec:discussion}

We study radial relaxation by varying the quartic coupling $\lambda_\Phi$ and the release width $\Delta$ while keeping the coupling ratio $u=y_K^2/\lambda_\Phi=2/5$. The Yukawa coupling therefore changes with the quartic according to $y_K=\sqrt{2\lambda_\Phi/5}$. We calculate the fluctuation amplification using the conservative background evolution with Hubble damping and treating fluctuations to linear order. The differences in amplification reflect changes in the PQ potential, the inflationary displacement and the duration of curvature release. The accompanying change in $y_K$ enters the quark mass and decay widths, with its effect on early energy transfer tested separately. At fixed $\lambda_\Phi$, the optimization over $u$ in Section~\ref{sec:decay} maximizes the vacuum fermion decay width.

Interactions that transfer energy from the coherent PQ condensate to the thermal bath can damp the radial oscillations. At fixed PQ potential and curvature-release history, this damping could extend our linear treatment to additional parameter values if it suppressed transverse fluctuation growth enough to keep the variance below the adopted cutoff. We test early energy loss in the plasma calculations of Appendix~\ref{app:micro_response} by varying the dimensionless damping normalization $C_{\rm diss}=0,1,10$ at the benchmark couplings. This changes the damping normalization and its associated thermal noise while retaining the same thermal force.

At the later decay epoch, the surviving saxion condensate transfers its energy to relativistic axions and, through the KSVZ quarks, to the visible bath. We calculate the resulting axion radiation abundance in Section~\ref{sec:results}. A cosmological history that satisfies the primordial isocurvature bound and retains a coherent condensate can therefore face an additional constraint from the radiation density. Interpreting that constraint as a bound on curvature release depends on the assumed potential and decay channels, since they determine both the surviving abundance and its conversion into radiation. The CMB comparison also involves two distinct components. With independent radial fluctuations suppressed during inflation and under the assumed single-clock reheating history, the decay radiation is adiabatic~\cite{Weinberg:2004kr}, while cold axion dark matter retains the angular isocurvature perturbation. A uniform PQ rotation leaves the radial decay source unchanged, consistent with this separation.

We stop the linear evolution when the variance of transverse PQ-field fluctuations reaches the adopted cutoff, $\mathcal V_{\rm band}=10^{-2}$. It is neither a CMB exclusion boundary nor evidence by itself for restoration of PQ symmetry. Establishing the outcome of trajectories that reach it involves the subsequent nonlinear evolution. Even below the cutoff, the accuracy of the radiation prediction depends on the accumulated backreaction on the condensate. This is particularly relevant to benchmark B, because fluctuations can affect a subdominant radial energy density while remaining negligible in the total cosmological density. The maximum fluctuation-energy ratio in Appendix~\ref{app:numerical_validation} measures their instantaneous importance, rather than an error bar on $\Delta N_{\rm eff}$.

At stronger Yukawa coupling, the quark mass, thermal force and dissipation change together. The positive thermal curvature can move the minimum towards the origin, while the fluctuations associated with thermal damping contribute to the angular variance. The outcome then depends on the competition between coherent energy loss, motion of the minimum and fluctuation growth~\cite{Mukaida:2012qn,Moroi:2013tea,Yokoyama:1998ju}. Studying this regime would require solving the dissipative field equations with independent variations of $y_K$ at fixed $\lambda_\Phi$.

In the weak Yukawa regime tested here, saxion decay becomes important only after radial relaxation, while the modeled plasma damping is too weak to prevent benchmark C from reaching the transverse fluctuation cutoff. Our decay estimates and plasma-response tests therefore provide no evidence that these interactions enlarge the parameter region that remains below the cutoff.

A Higgs portal, which we have taken to be negligible, could provide additional damping during radial relaxation, as studied in other axion cosmologies~\cite{Eroncel:2024rpe,Co:2017mop,Co:2020dya}. To suppress transverse amplification, this interaction would need to damp the radial oscillations before the fluctuations grow appreciably, while preserving the pre-inflationary PQ configuration. The portal would also modify the thermal potential and open additional saxion decay channels into visible particles. Establishing whether it enlarges the allowed parameter region would therefore require following these effects together with the accompanying thermal noise and the resulting dark-radiation abundance.

\section{Conclusions}
\label{sec:conclusions}

We have calculated the dark radiation produced by a curvature-displaced PQ condensate in a quartic KSVZ model. Following the radial field through the end of inflation determines the coherent saxion abundance that survives until decay. We use the oscillation action to determine the initial saxion density for the late fluid evolution. For one coloured KSVZ fermion and a negligible Higgs portal,
the small quartic coupling keeps perturbative decay slow during relaxation. The fermion mass threshold also limits visible energy transfer, keeping the axion branching fraction above approximately $64\%$ within the two-channel approximation.

The benchmark references A and B predict $\Delta N_{\rm eff}\simeq0.033$ and $0.12$, respectively. Their saxion energy fractions reach maxima of only $0.87\%$ and $3.1\%$, so both condensates decay during radiation domination. These results show how a subdominant saxion population can produce an appreciable radiation abundance. Once its oscillations become approximately harmonic, the condensate dilutes more slowly than the radiation bath, increasing its relative contribution before decay. The resulting axions remain relativistic to the present.

The main result is the strong dependence of this abundance on the duration of curvature release. At fixed $\lambda_\Phi=10^{-11}$,
increasing the release width from $\Delta=0.20$ to $0.50$ reduces $\Delta N_{\rm eff}$ from approximately $0.12$ to $0.0044$, a factor
of about $28$, while preserving the primordial angular power. The particle masses and decay widths are identical along this
sequence. The reduction therefore follows from the smaller radial excitation left when the field tracks its contracting minimum more closely. For the smooth reheating histories studied here, gradual release suppresses both transverse amplification and the coherent
energy available for late decay. The backreaction of transverse PQ-field fluctuations on the saxion condensate remains small in benchmark A and in cosmological histories with more gradual curvature release. For benchmark B, a nonlinear calculation would determine how much the growing fluctuations change the condensate energy left for decay.

Nonlinear lattice simulations would track the exchange of energy between the coherent saxion condensate and spatial fluctuations of the PQ field, including their feedback on the radial motion. They would determine the surviving saxion abundance and resolve any phase winding that develops~\cite{Kasuya:1997ha,Harigaya:2015hha,Ballesteros:2021bee,Buschmann:2021sdq,Benabou:2024msj,Gorghetto:2018myk,Gorghetto:2020qws,Buschmann:2019icd}. This would extend the abundance calculation into the region where fluctuation backreaction becomes appreciable.

The evolution in that region also depends on how reheating releases the PQ field from its inflationary minimum. A microscopic reheating model would determine the spacetime curvature and the formation of the thermal bath consistently, including curvature oscillations and changes in the relation between the bath pressure and energy density. With this thermal history specified, stronger Yukawa interactions could be studied by retaining the quark-mass and frequency dependence of the plasma response, together with the associated thermal fluctuations. Following these effects during the radial relaxation of the PQ condensate after inflation would establish whether the interactions preserve the pre-inflationary PQ configuration and how much coherent energy remains available for axion production through subsequent saxion decay.

\begin{acknowledgments}
AS thanks Peter W. Graham for introducing him to the topic and suggesting the idea for this work. We also thank him for reading the manuscript and sharing his perspective and feedback. AS is partially supported by the U.S. National Science Foundation (NSF) under Grant No.~PHY-2310363. AS also acknowledges support from a Universities Research Association Visiting Scholars Fellowship at Fermilab. We acknowledge the use of ChatGPT (OpenAI) for assistance in numerical scanning and sentence-level editing.
\end{acknowledgments}

\appendix

\setlength{\parindent}{1.5em}
\setlength{\parskip}{0pt}
\setlength{\emergencystretch}{2em}
\allowdisplaybreaks[1]

\section{KSVZ completion and decay assumptions}
\label{app:micro_decay}
\label{sec:portal_assumption}

The fermion channel contributes to visible radiation only if the KSVZ quarks produced in saxion decay can transfer their energy to Standard Model particles. We allow this through a PQ-preserving mixing interaction. The model contains one Dirac quark $\psi$ in the representation $(3,1,-1/3)$ of $SU(3)_c\times SU(2)_L\times U(1)_Y$ and the complex PQ scalar $\Phi$~\cite{Kim:1979if,Shifman:1979if}. With PQ charges $q_{\rm PQ}(\Phi,\psi_L,\psi_R)=(1,1,0)$ and vanishing charges for the Standard Model fields, the domain-wall number is one. These assignments also permit~\cite{DiLuzio:2017pfr}
\begin{equation}
 \mathcal L_{\rm mix}
 =-\epsilon_{\rm mix}\overline q_L^{\,(i)}\mathcal H\psi_R
 +\mathrm{h.c.},
 \label{eq:quarkmixing}
\end{equation}
where $q_L^{(i)}$ is a Standard Model quark doublet of generation $i$, $\mathcal H$ is the Higgs doublet and $\epsilon_{\rm mix}$ is the dimensionless mixing coupling. The heavy quark has vacuum mass $m_\psi=y_Kf_a/\sqrt2$. For $m_\psi\gg v$, where $v=246\,\mathrm{GeV}$ is the electroweak vacuum expectation value, summing the decay channels associated with the two Higgs-doublet components gives the total KSVZ-quark decay width
\begin{equation}
 \Gamma_\psi=\frac{|\epsilon_{\rm mix}|^2m_\psi}{16\pi}
 \left[1+\mathcal O\!\left(\frac{v^2}{m_\psi^2}\right)\right].
 \label{eq:quarkwidth}
\end{equation}

We choose $\epsilon_{\rm mix}=10^{-4}$, for which $\Gamma_\psi/\Gamma_s\simeq68$ in A and $203$ in B. Since the main saxion decay occurs near $H=\Gamma_s$, the daughter quarks decay much faster than the expansion at this epoch. This justifies treating the fermion channel as a source for the visible bath. The thermal quark population also decays efficiently before a substantial relic can survive, with $\Gamma_\psi/H\simeq220$--$700$ at $T=m_\psi/20$ across A--C. During the earlier radial relaxation, however, $\Gamma_\psi\ll H$, so quark decay does not promptly remove the particles produced at successive crossings of the PQ field through the origin.

The two-channel approximation also requires other saxion decay modes to be subleading. In particular, the symmetries allow a PQ--Higgs portal~\cite{Benabou:2023ghl},
\begin{equation}
 \Delta V_{\Phi H}
 =\lambda_{\Phi H}\left(|\Phi|^2-\frac{f_a^2}{2}\right)
 \mathcal H^\dagger\mathcal H,
 \label{eq:portal_potential}
\end{equation}
where $\lambda_{\Phi H}$ is the renormalized portal coupling. Expanding about the PQ vacuum gives the interaction $\lambda_{\Phi H}f_as\mathcal H^\dagger\mathcal H$, which permits saxion decay into Higgs pairs. For negligible Higgs masses, each of the four real Higgs components contributes a width $\lambda_{\Phi H}^2f_a^2/(32\pi m_s)$. Their sum therefore gives $\Gamma_{\mathcal H\mathcal H^\dagger}/\Gamma_{aa}^{(0)}=(\lambda_{\Phi H}/\lambda_\Phi)^2$. Our assumption $|\lambda_{\Phi H}|\ll0.1\lambda_\Phi$ makes this channel negligible in the decay calculation.

The same assumption keeps the Higgs contribution to the radial thermal force small. At the maximum benchmark temperature, $T/f_a=0.23$, the thermal curvature $\Pi_H=\lambda_{\Phi H}T^2/6$ satisfies $|\Pi_H|/(\lambda_\Phi f_a^2)<8.8\times10^{-4}$. Quark mixing generates a portal correction of order $N_cy_K^2|\epsilon_{\rm mix}|^2/(16\pi^2)$ per logarithmic scale interval, with $N_c=3$, which remains far below $0.1\lambda_\Phi$ for the chosen couplings. These estimates support neglecting the portal in both the radial evolution and the late decay. We assume that the scalar completion keeps the Higgs in its electroweak basin during inflation and that it enters the symmetric thermal state after reheating~\cite{Degrassi:2012ry,Elias-Miro:2012eoi,Hook:2014uia}.

Quark loops also permit saxion decay into gluons and hypercharge gauge bosons. The gluon width carries a suppression of order $\alpha_s^2/\pi^2$ relative to the axion width, with an analogous gauge suppression for the hypercharge channel~\cite{Mukaida:2012qn,Benabou:2023ghl}. The amplitudes must retain the finite quark-mass dependence for our open-fermion benchmarks. These channels remain subleading corrections to the two tree-level widths in the perturbative regime considered here.

Before the field settles near its vacuum, fermions can also be produced by the changing mass $m(t)=g_Y\overline X(t)$, where $g_Y=y_K/\sqrt2$ and $\overline X$ is the homogeneous Cartesian PQ field. As the field crosses the origin, the mass passes through zero and its variation can become nonadiabatic. For an isolated crossing at $t_{\rm cr}$ into initially empty states, neglecting thermal effects, the characteristic physical momentum is $p_{\rm prod}^2=g_Y|\dot{\overline X}(t_{\rm cr})|$, with occupation $n_p^{(1)}=\exp[-\pi p^2/p_{\rm prod}^2]$ per fermionic state of physical momentum $p$~\cite{Felder:1998vq,Greene:2000ew}.

The small quartic coupling limits the energy carried by this population relative to the condensate. For quartic oscillations of amplitude $A_X$, we use the frequency scale $\Omega_X=\sqrt{\lambda_\Phi}A_X$. At fixed $u=y_K^2/\lambda_\Phi$, rescaling momenta by $\Omega_X$ gives $\rho_\psi=N_c\Omega_X^4\mathcal I_\psi$, where $\mathcal I_\psi$ is the dimensionless energy integral over the occupied states, including spin and antiparticle multiplicities. Writing the condensate energy as $\rho_\Phi=\lambda_\Phi A_X^4\mathcal E_\Phi$, with dimensionless $\mathcal E_\Phi$, we obtain
\begin{equation}
 \frac{\rho_\psi}{\rho_\Phi}
 =N_c\lambda_\Phi\frac{\mathcal I_\psi}{\mathcal E_\Phi}.
 \label{eq:app_weak_yukawa_energy_scaling}
\end{equation}
Thus, over a finite interval of cold evolution at fixed dimensionless occupations, the fractional energy transfer is suppressed by $\lambda_\Phi$. In a thermal bath, the additional scale $T$ changes both the fermion occupations and the force on the PQ field. We treat those effects separately in Appendix~\ref{app:micro_thermal}.

\section{Thermal corrections, plasma response and additional populations}
\label{app:micro_thermal}
\label{app:micro_response}

\subsection{Thermal force and vacuum corrections}

The field-dependent KSVZ quark mass is $m_\psi(\rho)=y_K\rho/\sqrt2$, where $\rho=\sqrt2|\Phi|$ is the radial amplitude of the complex PQ field $\Phi$ and $y_K$ is the Yukawa coupling. At the vacuum, $\rho=f_a$, this becomes $m_\psi=y_Kf_a/\sqrt2$. If the quarks are in thermal equilibrium, this mass dependence gives a radial force through their free energy. We write the effective free-energy density as $\mathcal F=V_0+\Delta V_{\rm ren}+\Delta\mathcal F_\psi$, where $V_0$ is the tree potential in Eq.~\eqref{eq:model_potential}, $\Delta V_{\rm ren}$ is the renormalized vacuum correction and $\Delta\mathcal F_\psi$ is the thermal contribution. For a Dirac quark with $N_c=3$ colours at temperature $T$, the standard fermion thermal integral gives~\cite{Quiros:1999jp,Mukaida:2012qn,Laine:2016hma}
\begin{equation}
 \begin{aligned}
 \Delta\mathcal F_\psi
 &=-\frac{2N_cT^4}{\pi^2}\left[J_F(w)-J_F(0)\right],\\
 J_F(w)&=\int_0^\infty dv\,v^2
 \ln\left(1+e^{-\sqrt{v^2+w}}\right),
 \qquad w=\frac{m_\psi^2(\rho)}{T^2}.
 \end{aligned}
 \label{eq:full_fermion_thermal_function}
\end{equation}
Here $J_F(w)$ is the dimensionless fermionic thermal function, and $w=[m_\psi(\rho)/T]^2$ is the squared ratio of the field-dependent quark mass to the thermal bath temperature. The integration variable $v$ is momentum in units of $T$. The subtraction of $J_F(0)$ removes a field-independent term without changing the radial force. Keeping the full integral accounts for the decrease in the quark population when $m_\psi(\rho)$ becomes comparable to $T$.

For $m_\psi(\rho)\ll T$, the high-temperature expansion gives $\Delta\mathcal F_\psi=N_cy_K^2T^2\rho^2/24$~\cite{Quiros:1999jp}. Its positive curvature, $\Pi_\psi=N_cy_K^2T^2/12$, favours a smaller PQ radius and opposes the outward displacement caused by the Ricci scalar $R$. Adding this term to the tree potential gives the stationary radius $f_{R,T}$ and the squared mass at the origin,
\begin{equation}
 f_{R,T}^2=f_a^2+\frac{2\xi R-\Pi_\psi}{\lambda_\Phi},
 \qquad
 m_{\rm origin}^2=-\lambda_\Phi f_a^2-2\xi R+\Pi_\psi.
 \label{eq:thermal_radius_origin}
\end{equation}
Here $f_a$ is the vacuum radius, $\lambda_\Phi$ is the quartic coupling and $\xi$ is the curvature coupling. The quantity $m_{\rm origin}^2$ is the effective squared mass of small PQ-field fluctuations about $\rho=0$, determined by the curvature of the potential there. A negative value means that the origin is unstable. A broken minimum exists in this approximation when $f_{R,T}^2>0$. Since the thermal contribution depends only on $X^2+Y^2$, the same $\Pi_\psi$ must enter the homogeneous and transverse equations. The transverse squared mass $M_Y^2=\lambda_\Phi(\overline X^2-f_a^2)-2\xi R+\Pi_\psi$ then vanishes at the broken minimum, as required by PQ symmetry. For the benchmarks with $u=y_K^2/\lambda_\Phi=2/5$ and release width $\Delta=0.20$, we find $\Pi_\psi/(\lambda_\Phi f_a^2)\leq5.3\times10^{-3}$. Including this force in both Cartesian equations changes the peak transverse variance of B by $0.024\%$.

We also check that vacuum quark loops do not appreciably alter the assumed quartic potential. We write their Coleman--Weinberg contribution~\cite{PhysRevD.7.1888,Quiros:1999jp},
\begin{equation}
 \Delta V_{\rm CW}^{(\psi)}
 =-\frac{N_cm_\psi^4(\rho)}{16\pi^2}
 \left[\ln\frac{m_\psi^2(\rho)}{\mu^2}-\frac32\right],
 \qquad
 \epsilon_{\rm CW}=\frac{N_cy_K^4}{16\pi^2\lambda_\Phi}.
 \label{eq:ksvz_radiative_ratio}
\end{equation}
The scale $\mu$ is the renormalization scale. We choose finite counterterms so that the renormalized correction preserves the vacuum energy, radius and radial mass. The dimensionless parameter $\epsilon_{\rm CW}$ characterizes the size of the fermion Coleman--Weinberg correction relative to the tree-level quartic potential, with the logarithmic field dependence shown explicitly in Eq.~\eqref{eq:ksvz_radiative_ratio}. At fixed $u$, it becomes $3u^2\lambda_\Phi/(16\pi^2)$, which is small throughout the field excursion for the weak-Yukawa benchmarks. Since both the vacuum and thermal corrections are small, we use the tree-level potential to follow the early radial relaxation of the PQ condensate in these benchmarks.

\subsection{Local damping and thermal fluctuations}
\label{sec:reference_response}

The thermal force changes the position of the minimum, while collisions can remove energy from the moving PQ condensate. To test this second effect, we use the very-light-quark damping rate of Ref.~\cite{Mukaida:2012qn}, which scales as $g_Y^2\alpha_sT$ for radial Yukawa vertex $g_Y=y_K/\sqrt2$. We adopt the same local damping coefficient $\Upsilon$ for the Cartesian fields $\phi_i=(X,Y)$. The dimensionless coefficient $C_{\rm diss}$ specifies its normalization. We compare $C_{\rm diss}=0,1,10$ at strong coupling $\alpha_s=0.03$, retaining the same thermal force in all three calculations.

We turn on this response smoothly near the centre of the transition. With $N=\ln(a/a_e)$, where $a_e$ is the scale factor at that centre, the onset is $N_{\rm on}=0$ and its duration is $\delta N_{\rm on}=0.05$. Defining $q_N=(N-N_{\rm on})/\delta N_{\rm on}$, we use
\begin{equation}
 \begin{aligned}
 \Upsilon(N)&=W(N)\frac{C_{\rm diss}y_K^2\alpha_sT(N)}{2},\\
 W(N)&=
 \begin{cases}
 0,&q_N\leq0,\\
 3q_N^2-2q_N^3,&0<q_N<1,\\
 1,&q_N\geq1.
 \end{cases}
 \end{aligned}
 \label{eq:response_switch}
\end{equation}
The function $W$ is our prescribed onset of the bath response. Writing the conservative radial force as $\mathcal A\overline X$, with $\mathcal A=\lambda_\Phi(\overline X^2-f_a^2)-2\xi R+\Pi_\psi$, the homogeneous field and transverse Fourier mode $Y_k$ obey
\begin{align}
 \ddot{\overline X}+(3H+\Upsilon)\dot{\overline X}
 +\mathcal A\overline X&=0,\\
 \ddot Y_k+(3H+\Upsilon)\dot Y_k
 +\left(\frac{k^2}{a^2}+\mathcal A\right)Y_k&=\eta_k.
 \label{eq:app_transverse_response_equation}
\end{align}
Here $H$ is the Hubble rate, $k$ is comoving momentum, and $\eta_k$ is the random force exerted by the bath. Applying the same damping to both Cartesian fields preserves the uniform-rotation solution $Y_0=\delta\theta_0\overline X$ for a constant angle $\delta\theta_0$ when $\eta_k=0$.

The random force must accompany collisional damping because the bath both absorbs coherent energy and excites fluctuations~\cite{Mukaida:2012qn,Moroi:2013tea,Moroi:2014mqa,Berera:2008ar}. In the classical local limit, the fluctuation--dissipation relation fixes the correlation of the zero-mean Cartesian forces~\cite{Berera:2008ar},
\begin{equation}
 \langle\eta_i(t,\boldsymbol x)\eta_j(t',\boldsymbol x')\rangle
 =2T\Upsilon\,\delta_{ij}a^{-3}
 \delta(t-t')\delta^{(3)}(\boldsymbol x-\boldsymbol x').
 \label{eq:app_local_pq_noise}
\end{equation}
The brackets denote an ensemble average, $\boldsymbol x$ is comoving position and $\delta_{ij}$ is the unit tensor in field space. The factor $a^{-3}$ converts the physical spatial delta function to comoving coordinates. The noise strength is therefore fixed by the same $\Upsilon$ that damps the field motion.

We calculate the variance generated by this force by evolving the bath covariance spectra. These are the field spectrum $P_{YY}$, the velocity spectrum $P_{vv}$ with $v_k=\dot Y_k$, and their real symmetrized cross spectrum $P_{Yv}$; each definition factors out $(2\pi)^3\delta^{(3)}(\boldsymbol k-\boldsymbol k')$. Differentiating these covariances using the field equation and noise correlation gives
\begin{equation}
 \begin{aligned}
 \dot P_{YY}&=2P_{Yv},\\
 \dot P_{Yv}&=P_{vv}-\omega_{Y,k}^2P_{YY}-\gamma_YP_{Yv},\\
 \dot P_{vv}&=-2\omega_{Y,k}^2P_{Yv}-2\gamma_YP_{vv}
 +\frac{2T\Upsilon}{a^3}.
 \end{aligned}
 \label{eq:noise_covariance_system}
\end{equation}
Here $\gamma_Y=3H+\Upsilon$ is the total damping coefficient and $\omega_{Y,k}^2=k^2/a^2+\mathcal A$ is the effective squared mode frequency. We set the bath covariances to zero at $N_{\rm on}$, so this calculation follows the fluctuations generated after the response begins. The noise first sources the velocity variance and subsequently the field variance. Integrating over the resolved momentum interval gives
\begin{equation}
 \mathcal V_{\rm bath}
 =\frac{1}{\pi^2f_a^2}
 \int_{k_{\min}}^{k_{\max}}\frac{dk\,k^2}{2\pi^2}P_{YY},
 \qquad
 \mathcal V_{\rm tot}=\mathcal V_{\rm inf}+\mathcal V_{\rm bath}.
 \label{eq:noise_variance_split}
\end{equation}
The limits $k_{\min}$ and $k_{\max}$ are the smallest and largest included comoving momenta. The damped inflationary contribution $\mathcal V_{\rm inf}$ and the bath contribution are statistically independent, so their variances add.

Since the PQ field is a spectator, we obtain the temperature from the prescribed reheating history. During curvature release we use $H/H_*=(1+e^{N/\Delta})^{-2\Delta}$, the leading expression in the small slow-roll correction, where $H_*$ is the inflationary Hubble normalization and $\Delta$ is the release width. We include 160 logarithmically spaced modes over $0.1\leq\kappa\leq10$, with $\kappa=k/(a_eH_e)$ and $H_e=H(0)$. The initial scale factor $a_i$ and Hubble rate $H_i$ satisfy $k_{\min}/(a_iH_i)=60$, and the relative integration tolerance is $3\times10^{-10}$.

A local damping term requires the plasma to relax faster than the field and background evolve~\cite{Mukaida:2012qn,Berera:2008ar}. We check this using the plasma rate $\alpha_sT$ and a positive estimate of the field frequency, $\Omega_{\rm ctrl}^2=\lambda_\Phi(3\overline X^2+f_a^2)+2\xi R+\Pi_\psi$. Throughout the response interval, $\Omega_{\rm ctrl}/(\alpha_sT)\lesssim0.054$, $m_\psi(\rho)/(\alpha_sT)<0.013$ and $k/(a\alpha_sT)<0.022$. Expansion and the changing quark mass are also slow, with $H/(\alpha_sT)<0.0022$ and $g_Y|\dot{\overline X}|/(\alpha_sT)^2<1.7\times10^{-5}$. These checks support the local and very-light-quark approximations over the interval used.

Even for $C_{\rm diss}=10$, the accumulated damping $\mathcal D_\psi=\int\Upsilon\,dt$ is only $1.71\times10^{-6}$ in A and $9.39\times10^{-7}$ in B. The bath contributes less than $0.1\%$ of the total transverse variance at its respective maxima. Damping slightly reduces the inflationary contribution, while thermal noise produces a small net increase in the total. Benchmark C reaches $\mathcal V_{\rm tot}=10^{-2}$ near $N=3.296$ for all three normalizations. These conclusions remain unchanged when we repeat the calculation with more momentum modes and an earlier onset of plasma damping and thermal noise, at $N_{\rm on}=-0.2$. Thus the adopted plasma response and initial bath covariance have little effect on the benchmark relaxation trajectories and do not prevent C from reaching the cutoff.

\subsection{Thermal populations and axion scattering}
\label{sec:benchmark_thermal_estimates}

The thermal bath can also produce axions and saxions independently of the coherent condensate. For axions, the relevant high-temperature processes include $\psi g\leftrightarrow\psi a_{\rm ax}$ and $\psi\bar\psi\leftrightarrow ga_{\rm ax}$, where $g$ is a gluon and $a_{\rm ax}$ is the axion field~\cite{DEramo:2021lgb}. The radial Yukawa interaction permits the corresponding saxion processes. We estimate these populations in the freeze-in limit, where their abundances remain far below equilibrium and reactions that deplete the produced populations can be neglected~\cite{Hall:2009bx}. We write the axion energy-production rate $Q_a^{\rm th}$ and saxion number-production rate $\gamma_s^{\rm th}$ per unit physical volume as
\begin{equation}
 Q_a^{\rm th}=c_a y_K^2\alpha_sT^5,
 \qquad
 \gamma_s^{\rm th}=c_s y_K^2\alpha_sT^4.
 \label{eq:thermal_production_scaling}
\end{equation}
The dimensionless coefficients $c_a$ and $c_s$ contain the thermal collision integrals. We leave them explicit because these expressions estimate the production rates from their coupling and temperature dependence.

During radiation domination, $H=c_RT^2/M_{\rm P}$ with $c_R=\sqrt{\pi^2g_*/90}$, where $M_{\rm P}$ is the reduced Planck mass and $g_*$ counts energy degrees of freedom. To remove the dilution caused by expansion, we use the axion energy ratio $R_a^{\rm th}=\rho_a^{\rm th}/\rho_v$ and the saxion yield $Y_s^{\rm th}=n_s^{\rm th}/s_v$. Here $\rho_v$ and $s_v$ are the visible energy and entropy densities, while $\rho_a^{\rm th}$ and $n_s^{\rm th}$ describe the produced axions and saxions. We integrate the axion energy-production and saxion number-production rates while keeping the effective energy and entropy degrees of freedom, $g_*$ and $g_{*s}$, constant. For this estimate, we neglect further axion production once the temperature falls below $m_\psi$ and further saxion production below $m_s$. Starting at the maximum thermal bath temperature $T_{\max}$ then gives
\begin{equation}
 \begin{aligned}
 R_a^{\rm th}
 &\simeq\frac{30c_a y_K^2\alpha_sM_{\rm P}}{\pi^2g_*c_R}
 \left(\frac{1}{m_\psi}-\frac{1}{T_{\max}}\right),\\
 Y_s^{\rm th}
 &\simeq\frac{45c_s y_K^2\alpha_sM_{\rm P}}{2\pi^2g_{*s}c_R}
 \left(\frac{1}{m_s}-\frac{1}{T_{\max}}\right).
 \end{aligned}
 \label{eq:thermal_production_estimate}
\end{equation}
The inverse-temperature dependence shows that most production occurs near the assumed mass thresholds. For $g_*=g_{*s}=117.25$ and $\alpha_s=0.03$, we obtain $R_a^{\rm th}/c_a=2.6\times10^{-5}$ in the benchmark A and $1.5\times10^{-5}$ in the benchmark B, while $Y_s^{\rm th}/c_s=6.1\times10^{-6}$ and $3.5\times10^{-6}$, respectively.

After the produced saxions become nonrelativistic, their energy dilutes more slowly than radiation. Before decay, their energy fraction therefore grows approximately as $(4/3)(m_s/T)Y_s^{\rm th}$ for equal energy and entropy degree counts. At the decay temperature $T_\Gamma\equiv T(H=\Gamma_s)$, this gives $2.8\times10^{-4}c_s$ ib A and $2.1\times10^{-4}c_s$ in B; saxion decay reduces these fractions further. For coefficients of order unity, these thermally produced populations give small corrections to the coherent sources in both benchmarks.

The open fermion channel also allows thermal quarks to produce saxions through $\psi\bar\psi\to s$. By detailed balance, this inverse-decay source is determined by the vacuum width $\Gamma_{\psi\bar\psi}^{(0)}$. Using classical Boltzmann occupations and integrating the source in radiation domination, as in the freeze-in calculation of Ref.~\cite{Hall:2009bx}, we obtain
\begin{equation}
 Y_s^{\rm ID}\simeq
 \frac{135}{8\pi^3g_{*s}c_R}
 \frac{\Gamma_{\psi\bar\psi}^{(0)}M_{\rm P}}{m_s^2}.
 \label{eq:thermal_inverse_decay}
\end{equation}
The superscript ID denotes inverse decay. The yields are $1.35\times10^{-6}$ in A and $7.82\times10^{-7}$ in B, corresponding to energy fractions of $6.2\times10^{-5}$ and $4.7\times10^{-5}$ at $T_\Gamma$ before the produced saxions decay. These contributions are also small compared with the benchmark condensates.

Finally, we check whether scattering on the bath could thermalize the axions from condensate decay. At temperatures below the KSVZ-quark mass, the axion--gluon interaction controls the rate~\cite{DEramo:2021lgb}. For axion energy $E_a$, its dependence on the coupling and temperature gives the estimate
\begin{equation}
 \frac{\Gamma_a^{\rm scatt}}{H}
 \sim C_{\rm scatt}(E_a/T)
 \frac{\alpha_s^3M_{\rm P}T}{c_Rf_a^2}.
 \label{eq:late_axion_scattering}
\end{equation}
Here $\Gamma_a^{\rm scatt}$ is the scattering rate and $C_{\rm scatt}$ is a dimensionless function containing the momentum dependence and plasma-screening effects~\cite{Bouzoud:2024bom}. The explicit prefactor is approximately $2\times10^{-8}$ at $T\sim10^7\,\mathrm{GeV}$, $f_a=10^{14}\,\mathrm{GeV}$ and $\alpha_s=0.03$, with injection energies $E_a/T\simeq17$ to $23$. For perturbative values of $C_{\rm scatt}$, scattering of the decay axions on thermal gluons and Standard Model quarks is much slower than the expansion, so the axions do not thermalize with the visible bath. The quoted $\Delta N_{\rm eff}$ therefore describes freely propagating axions from the coherent condensate, with the additional thermal populations estimated separately above.

\section{Background construction and initial conditions}
\label{app:background_methods}

We reconstruct a canonical, minimally coupled inflaton $\sigma$ with potential $V_\sigma$ that reproduces the expansion rate and radiation fraction specified in Section~\ref{sec:numerics}. The inflaton transfers energy to radiation at a rate $Q$ per unit physical volume and cosmic time. This construction relates the decrease in spacetime curvature to the growth of the thermal bath, so both enter the PQ-field evolution consistently. We neglect the small PQ contribution to the gravitational equations and treat radiation as an ideal fluid with pressure $p_r=\rho_r/3$. We retain the convention that a prime denotes $d/dN$ and a dot denotes a cosmic-time derivative.

The Raychaudhuri equation for the inflaton and radiation is $-2M_{\rm P}^2\dot H=\dot\sigma^2+4\rho_r/3$~\cite{Kinney:1997ne}. Using the prescribed $\epsilon_H$ and radiation fraction $r_r$, we obtain
\begin{equation}
 \sigma'=M_{\rm P}\sqrt{2(\epsilon_H-2r_r)}.
 \label{eq:num_reconstructed_trajectory}
\end{equation}
We choose the solution with increasing $\sigma$. The functions defined in Section~\ref{sec:numerics} satisfy $\epsilon_H-2r_r=(1-L)(2L+\epsilon_{\rm sr})>0$ during the transition, ensuring positive inflaton kinetic energy. Since $\sigma$ evolves monotonically, its potential can be specified along the trajectory. Subtracting the kinetic and radiation densities from the Friedmann constraint gives
\begin{equation}
 V_\sigma(N)=M_{\rm P}^2H^2(3-\epsilon_H-r_r).
 \label{eq:num_reconstructed_potential}
\end{equation}
For the small slow-roll normalization used here, $V_\sigma$ remains positive and approaches zero as radiation comes to dominate. These equations therefore define a canonical inflaton potential over the field range needed for the calculation.

The radiation continuity equation, $\dot\rho_r+4H\rho_r=Q$, determines the required energy transfer~\cite{Giudice:2000ex}. Differentiating the prescribed radiation density gives
\begin{equation}
 \frac{Q}{H}=3M_{\rm P}^2H^2
 \left[r_r'+(4-2\epsilon_H)r_r\right].
 \label{eq:num_reconstructed_source}
\end{equation}
The source is positive for the histories considered here, and the inflaton loses the energy received by radiation, preserving total energy conservation. This reconstruction specifies the background energy transfer without selecting a microscopic interaction. We additionally assume that its microscopic realization has a single cosmological clock. On superhorizon scales, different regions then follow locally time-shifted copies of the same inflaton and radiation trajectory, as in the separate-universe description of adiabatic perturbations~\cite{Wands:2000dp}.

We normalize the expansion at horizon exit of the CMB reference mode $k_*$. With $N_*>0$ denoting the interval from horizon exit to $N=0$, we have $H(-N_*)=H_*$ and define $L_*=L(-N_*)$. Integrating the prescribed $\epsilon_H$ gives
\begin{equation}
 \frac{H(N)}{H_*}
 =\exp\left\{
 -2\Delta\ln\frac{1+e^{N/\Delta}}{1+e^{-N_*/\Delta}}
 -\frac{\epsilon_0}{b}
 \left[L^{b\Delta}-L_*^{b\Delta}\right]
 \right\}.
 \label{eq:num_reconstructed_H}
\end{equation}
The first term gives the late radiation-era behaviour $H\propto a^{-2}$, while the second retains the slow variation during inflation. The curvature $R=6(2-\epsilon_H)H^2$ follows from the same history and vanishes as $\epsilon_H$ approaches two.

At horizon exit, radiation and inflaton energy transfer are negligible. The leading slow-roll scalar spectrum therefore fixes $\epsilon_{H,*}\equiv\epsilon_H(-N_*)=H_*^2/(8\pi^2M_{\rm P}^2A_s)$~\cite{Kinney:1997ne}, with the normalization $A_s=2.1\times10^{-9}$ adopted in the main text~\cite{Planck:2018vyg}. Evaluating $\epsilon_H$ at this epoch gives
\begin{equation}
 \epsilon_0=
 \frac{\epsilon_{H,*}-2L_*}
 {(1-L_*)L_*^{b\Delta}}.
 \label{eq:num_scalar_normalization}
\end{equation}
Thus the scalar amplitude determines $\epsilon_0$ once $H_*$ and $N_*$ are specified.

We determine $N_*$ by matching the reference wavelength through the subsequent thermal history~\cite{Liddle:2003as}. Assuming that gauge interactions establish an equilibrium bath during reheating~\cite{Harigaya:2013vwa}, the relation $\rho_r=\pi^2g_*T^4/30$~\cite{Giudice:2000ex} gives
\begin{equation}
 T(N)=
 \left[\frac{90M_{\rm P}^2H^2r_r}{\pi^2g_*}\right]^{1/4}.
 \label{eq:num_bath_temperature}
\end{equation}
During the early relativistic stage, the Standard Model particles and one Dirac KSVZ quark give $g_*=g_{*s}=117.25$. The thermal and curvature forces on the PQ field therefore follow from the same background.

We define the radiation-era normalization by $H_\infty=\lim_{N\to\infty}e^{2N}H(N)$ and $T_\infty=[90M_{\rm P}^2H_\infty^2/(\pi^2g_*)]^{1/4}$. After the transition, $H\simeq H_\infty e^{-2N}$ and $T\simeq T_\infty e^{-N}$ while the relativistic particle content is fixed. These constants normalize the early radiation era rather than the actual expansion at arbitrarily late times. For this matching step, we neglect the small entropy injection from subsequent saxion decay. Conservation of visible comoving entropy, $g_{*s}T^3a^3=\mathrm{constant}$, then relates this stage to the present scale factor $a_0$ and photon temperature $T_{\gamma,0}=2.7255\,\mathrm K$~\cite{Fixsen_2009}. Combining this relation with $k_*=a_e e^{-N_*}H_*$ gives
\begin{equation}
 N_*=\ln\left[
 \frac{H_*}{k_*/a_0}
 \frac{T_{\gamma,0}}{T_\infty}
 \left(\frac{g_{*s,0}}{g_{*s}}\right)^{1/3}
 \right],
 \qquad g_{*s,0}=3.909.
 \label{eq:num_reference_matching}
\end{equation}
Here $k_*/a_0$ is the present physical wavenumber, and $g_{*s,0}$ is the present visible entropy count relative to the photon temperature, including the lower neutrino temperature. The cube-root factor accounts for entropy redistribution as massive particles disappear. We solve Eqs.~\eqref{eq:num_scalar_normalization} and \eqref{eq:num_reference_matching} together at each release width. Appendix~\ref{app:numerical_validation} evaluates the correction from saxion entropy production and includes the temperature-dependent KSVZ-quark pressure and density in the late fluid evolution. The early reconstruction instead uses ideal radiation with $\rho_r-3p_r=0$.

We initialize the homogeneous PQ field at $N=-12$, before curvature release, with $\overline X=f_R$, $\dot{\overline X}=\dot f_R$ and $\overline Y=0$. During nearly de Sitter inflation, the radial mass at the displaced minimum satisfies $m_{s,\min}^2=2\lambda_\Phi f_R^2\simeq48\xi H^2$ for $f_R\gg f_a$. With $\xi=1$, the radial response is faster than expansion, supporting tracking of the slowly moving minimum. Writing the homogeneous displacement as $\delta X_{\rm tr}=\overline X-f_R$ and linearizing Eq.~\eqref{eq:model_background} gives
\begin{equation}
 \ddot{\delta X}_{\rm tr}+3H\dot{\delta X}_{\rm tr}
 +m_{s,\min}^2\delta X_{\rm tr}
 =-(\ddot f_R+3H\dot f_R).
 \label{eq:model_tracking}
\end{equation}
The displacement $\delta X_{\rm tr}$ is distinct from the spatial fluctuation $\delta X=X-\overline X$ considered below. The right-hand side describes the driving caused by the moving minimum. A gradual curvature release reduces this driving relative to the restoring force in the histories studied here, leaving smaller radial oscillations. This linearized equation describes only small departures from the minimum; the numerical evolution uses the full signed-field equation, including crossings of the origin.

Each transverse mode $y_k$ is initialized at an early time $t_i$ when $k/(a_iH_i)\gg1$, with $a_i=a(t_i)$ and $H_i=H(t_i)$. We also require momentum to dominate the effective transverse mass and the frequency to vary slowly over an oscillation. The canonical mode $a y_k$ then approaches the positive-frequency vacuum solution $e^{-ik\tau}/\sqrt{2k}$~\cite{Kinney:1997ne}, where $d\tau=dt/a$ defines conformal time. Differentiating this solution gives
\begin{equation}
 y_k(t_i)=\frac{e^{-ik\tau_i}}{a_i\sqrt{2k}},
 \qquad
 \dot y_k(t_i)=
 \left(-H_i-i\frac{k}{a_i}\right)y_k(t_i),
 \label{eq:num_BD}
\end{equation}
with $\tau_i=\tau(t_i)$. An overall constant complex phase leaves the power spectrum unchanged.

Finally, we check the spectator approximation. The nonminimal coupling makes the coefficient of $R/2$ equal to $F(\overline X)=M_{\rm P}^2+2\xi\overline X^2$~\cite{Rigouzzo:2025ycb}. Its variation contributes $-3H\dot F=-12\xi H\overline X\dot{\overline X}$ to the homogeneous gravitational equation. To estimate the neglected PQ terms without cancellations, we define
\begin{equation}
 \mathcal E_{\rm PQ}=
 \frac12\dot{\overline X}^{\,2}
 +\frac{\lambda_\Phi}{4}(\overline X^2-f_a^2)^2
 +|\xi R\overline X^2|
 +12|\xi H\overline X\dot{\overline X}|.
 \label{eq:model_spectator_energy}
\end{equation}
This positive diagnostic includes the homogeneous kinetic and vacuum potential energies and the magnitudes of the curvature and gravitational derivative terms. It is not a separately conserved PQ energy density. We compare it with the background density and monitor the fractional change in the effective squared Planck mass through
\begin{equation}
 \epsilon_{\rm PQ}=
 \frac{\mathcal E_{\rm PQ}}{3M_{\rm P}^2H^2},
 \qquad
 \epsilon_{\rm grav}=
 \frac{2\xi\overline X^2}{M_{\rm P}^2}.
 \label{eq:model_spectator}
\end{equation}
Across the release widths and quartic couplings considered here, the respective maxima are $3.19\times10^{-5}$ and $8.75\times10^{-6}$. These small values support the spectator treatment throughout the resolved evolution.

\section{Field evolution and the late axion abundance}
\label{app:numerical_validation}

\subsection{Fluctuation spectra}

Radial oscillations amplify transverse PQ-field fluctuations, which become angular axion waves after the background settles near a broken minimum. We calculate their spectra and energy relative to the surviving condensate. Using $\kappa=k/(a_eH_e)$ and defining the dimensionless mode amplitude $z_\kappa=k^{3/2}y_k/f_a$, Eq.~\eqref{eq:model_variance} becomes
\begin{equation}
 \mathcal V_{\rm band}=\frac{1}{2\pi^4}
 \int d\ln\kappa\,|z_\kappa|^2.
 \label{eq:app_mode_normalization}
\end{equation}
The integral covers $0.1\leq\kappa\leq10$, with 120 momentum modes for benchmarks A and C and 480 for B.\footnote{We integrate the background and fluctuation equations using an adaptive Runge--Kutta method, with a relative tolerance of $2\times10^{-10}$.} For each logarithmic group of modes, we impose Eq.~\eqref{eq:num_BD} when the smallest momentum satisfies $k/(a_iH_i)=80$, placing all modes in the group well inside the Hubble radius. The reported variance includes the vacuum contribution within the resolved interval and is not a vacuum-subtracted particle abundance. The short-wavelength vacuum variance per logarithmic momentum interval is $\Delta_{Y,\rm vac}^2=k^2/(4\pi^2a^2)$.

\begin{figure}[tbp]
 \centering
 \includegraphics[width=\textwidth]{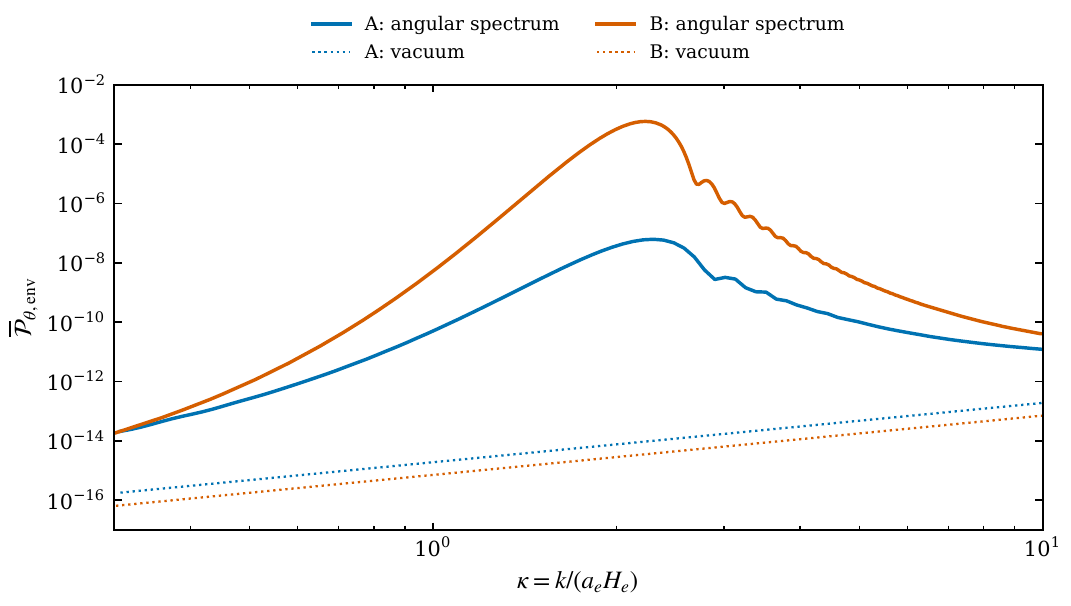}
 \caption{Angular-wave spectra after radial settling for benchmark A (blue) at $N_f=3.91$ and benchmark B (orange) at $N_f=4.40268$. Solid curves show Eq.~\eqref{eq:app_averaged_angular_envelope}, which includes the angular field and its time derivative and averages over the final $\delta N=0.15$ e-fold. Both axes are logarithmic, with $0.3\leq\kappa\leq10$ displayed. Dotted curves show the vacuum power $k^2/(4\pi^2a_f^2f_a^2)$, where $a_f=a(N_f)$. The vacuum curve for B lies lower because it is evaluated later and redshifts as $a_f^{-2}$. These waves arise during radial relaxation; subsequent condensate decay produces the radiation shown in Fig.~\ref{fig:late_energy_transfer}.}
 \label{fig:angular_spectra}
\end{figure}

After the background settles near one broken minimum and no longer crosses the origin, the linear angular perturbation is $\vartheta_k=y_k/\overline X$. Its derivative, $\dot\vartheta_k=(\dot y_k-\dot{\overline X}y_k/\overline X)/\overline X$, separates angular motion from the changing radius. With physical momentum $p=k/a$, the polar kinetic term gives the angular-wave energy density within the resolved band,
\begin{equation}
 \rho_{\theta,\rm band}=\int d\ln k\,
 \frac{k^3\overline X^2}{4\pi^2}
 \left(|\dot\vartheta_k|^2+p^2|\vartheta_k|^2\right).
 \label{eq:app_angular_wave_energy}
\end{equation}
The QCD potential is negligible at this epoch. Retaining both angular kinetic and gradient energies reduces sensitivity to the wave phase, motivating the envelope spectrum
\begin{equation}
 \mathcal P_{\theta,\rm env}(k)=\frac{k^3}{4\pi^2}
 \left(|\vartheta_k|^2+\frac{|\dot\vartheta_k|^2}{p^2}\right).
 \label{eq:app_angular_envelope}
\end{equation}
This spectrum satisfies $d\rho_{\theta,\rm band}/d\ln k=\overline X^2p^2\mathcal P_{\theta,\rm env}$. To compare the waves at a common final epoch $N_f$, we average over the preceding $\delta N=0.15$ e-fold. With $x=\overline X/f_a$, the spectrum in Fig.~\ref{fig:angular_spectra} is
\begin{equation}
 \overline{\mathcal P}_{\theta,\rm env}(k;N_f)
 =\frac{1}{\delta N}\int_{N_f-\delta N}^{N_f}dN\,
 x^2(N)e^{2(N-N_f)}\mathcal P_{\theta,\rm env}(k;N).
 \label{eq:app_averaged_angular_envelope}
\end{equation}
The factor $x^2$ converts the instantaneous radius to the fixed normalization $f_a$. After significant amplification ends, freely propagating subhorizon waves have energy density proportional to $a^{-4}$, while $p^2\propto a^{-2}$. The normalized spectrum therefore redshifts as $a^{-2}$, giving the factor $e^{2(N-N_f)}$.

The strongest amplification occurs near $\kappa\sim1$. At the final epoch in B, the angular waves carry $1.34\times10^{-9}$ of the visible radiation energy and $1.08\times10^{-3}$ of the coherent saxion energy. The surviving condensate therefore supplies the dominant axion radiation through its later decay.

We verify two exact properties of Eq.~\eqref{eq:model_modes}: the canonical Wronskian $a^3(y_k\dot y_k^*-y_k^*\dot y_k)=i$, which checks quantum normalization, and the uniform-rotation solution $y_0=\delta\theta_0\overline X$ for constant angle $\delta\theta_0$. Here the star denotes complex conjugation. For B, increasing the number of modes from 240 to 480 changes the final variance by $0.085\%$, while the integrated spectral difference between 120 and 480 modes is $0.38\%$. Extending the interval to $0.01\leq\kappa\leq30$ changes the peak dimensionless variance by less than $4\times10^{-13}$ in absolute terms.

For Fig.~\ref{fig:release_robustness}, we use 80 modes initialized at $k/(a_iH_i)=60$. We repeat six histories near the cutoff $\mathcal V_{\rm band}=10^{-2}$ with 240 to 320 modes initialized at $k/(a_iH_i)=80$. The peak variance changes by at most $0.038\%$ for trajectories below the cutoff, and the cutoff-crossing epoch changes by at most $2.6\times10^{-6}$ e-fold for those that reach it. Enlarging the momentum interval preserves these classifications. These tests establish numerical stability of the linear evolution; the physical effect of the fluctuations on the condensate requires a separate assessment.

\subsection{Backreaction and the surviving condensate}
\label{app:revision_matching}

The oscillating background can also amplify radial PQ-field fluctuations. Writing $\delta X=X-\overline X$ and denoting its Fourier amplitude by $x_k$, expansion of Eq.~\eqref{eq:model_potential} to quadratic order gives
\begin{equation}
 \ddot x_k+3H\dot x_k+
 \left[\frac{k^2}{a^2}+\lambda_\Phi(3\overline X^2-f_a^2)-2\xi R\right]x_k=0.
 \label{eq:app_longitudinal_mode}
\end{equation}
The time-dependent frequency can produce tachyonic growth or parametric resonance. We initialize the radial modes in the adiabatic vacuum at $p/H=80$, retaining both momentum and radial mass in the frequency; expansion corrections are of relative order $H^2/p^2$. Using 160 modes, we find quadratic radial fluctuation energies of $2.3\times10^{-7}$ and $8.7\times10^{-5}$ of the condensate energy at matching in A and B, respectively. For $\Delta=0.30,0.40,0.50$, the dimensionless radial variance $\langle(\delta X)^2\rangle_{\rm band}/(\pi^2f_a^2)$ remains below $1.3\times10^{-11}$ through $N=4.8$, and the radial fluctuation energy remains small over these intervals.

To assess feedback on the homogeneous field, we define $\sigma_X^2=\langle(\delta X)^2\rangle_{\rm band}$ and $\sigma_Y^2=\langle Y^2\rangle_{\rm band}$, where the brackets denote an ensemble average over the resolved momentum interval. Averaging the quartic radial force to quadratic order gives the leading Hartree correction $\delta F_X=\lambda_\Phi\overline X(3\sigma_X^2+\sigma_Y^2)$. Its importance depends on the condensate dynamics, even when the PQ sector is negligible in the total cosmological density. We also define
\begin{equation}
 \mathcal E_Y^+=\frac12\left\langle\dot Y^2+\frac{|\nabla Y|^2}{a^2}
 +\lambda_\Phi\overline X^2Y^2\right\rangle_{\rm band},
 \qquad \mathcal B_Y=\frac{\mathcal E_Y^+}{\rho_s}.
 \label{eq:positive_fluctuation_diagnostic}
\end{equation}
Here $\nabla$ differentiates with respect to comoving coordinates, and $\rho_s$ is the homogeneous kinetic plus vacuum potential energy density. The positive diagnostic $\mathcal E_Y^+$ retains the transverse kinetic, gradient and quartic interaction terms. It is not the complete quadratic fluctuation energy, which also contains potentially negative terms. The ratio $\mathcal B_Y$ measures the retained contributions relative to the condensate.

Over $0\leq N\leq4.5$, the maximum $\mathcal B_Y$ is $1.5\times10^{-5}$ in A and $0.086$ in B, and it remains below $3.2\times10^{-7}$ for the gradual-release sequence. These small contributions support the homogeneous calculation for A and the more gradual releases. In B, determining the accumulated effect of transverse fluctuations on the surviving condensate requires nonlinear evolution and remains the principal physical uncertainty. The maximum $\mathcal B_Y$ measures instantaneous contributions, not the fractional uncertainty in $\Delta N_{\rm eff}$, which depends on the entire relaxation trajectory.

For Fig.~\ref{fig:release_robustness}, we compare the surviving condensates at a common small oscillation amplitude. The matching epoch $N_m$ is the first epoch at $N\geq2.5$ when a trajectory below the transverse cutoff satisfies
\begin{equation}
 \mathcal A_E(N_m)\equiv\frac{\sqrt{2\rho_s(N_m)}}{m_sf_a}=0.04,
 \qquad
 \rho_s=\frac12(Hf_ax')^2+\frac{\lambda_\Phi f_a^4}{4}(x^2-1)^2.
 \label{eq:robustness_matching}
\end{equation}
The quantity $\mathcal A_E$ is the amplitude inferred from the energy, in units of $f_a$. The dimensionless comoving energy $\mathcal C_s=e^{3N_m}\rho_s(N_m)/(\lambda_\Phi f_a^4)$ removes the leading matter-like dilution~\cite{PhysRevD.28.1243,Masso:2005zg}. At matching, $2\xi R/(\lambda_\Phi f_a^2)<3.5\times10^{-6}$, so the curvature correction is small. Changing the amplitude condition to $0.03$ or $0.05$ changes $\mathcal C_s$ by less than $0.2\%$ in A and B.

For the late decay calculation, we use the oscillation action $J$ defined in Eq.~\eqref{eq:main_action_matching}. Once the curvature and thermal forces are negligible, $a^3J$ is conserved to adiabatic order when expansion is slow over a cycle and saxion decay removes negligible energy~\cite{Masso:2005zg}. Defining the dimensionless energy $e=\rho_s/(m_s^2f_a^2)$ and action $j=J/(m_sf_a^2)$, the quartic potential gives
\begin{equation}
 j(e)=\frac1\pi\int_{x_-}^{x_+}dx\,
 \sqrt{2\left[e-\frac{(x^2-1)^2}{8}\right]},
 \qquad x_\pm=\sqrt{1\pm\sqrt{8e}}.
 \label{eq:revised_action}
\end{equation}
The turning points $x_\pm$ bound an orbit in the positive well for $0<e<1/8$, below the central barrier; reflection gives the negative-well orbit. In the harmonic limit, $j\simeq e$, so the condensate energy density approaches $m_sJ$. Averaging over six complete cycles gives $J_m=m_sf_a^2\langle a^3j\rangle_{\rm cycles}/a_m^3$, where $a_m=a(N_m)$ and the brackets denote the cycle average. This prescription reduces the oscillation-phase dependence of the initial fluid density.

At the matching epochs for A and B, the amplitudes are approximately $0.04f_a$ and $m_s/H>10^3$. Moving the matching epoch across one e-fold changes $\Delta N_{\rm eff}$ by less than $2.3\times10^{-6}$ relatively. For $\Delta=0.20,0.30,0.40,0.50$ in Table~\ref{tab:release_dark_radiation}, we use $N_m=4.8,4.8,5.0,5.3$, respectively, allowing the more gradual releases to settle later. Shifting these epochs by $0.25$ e-fold changes the abundances by less than $1.5\times10^{-5}$ relatively.

\subsection{Late decay and axion radiation}

At matching, we initialize $\rho_{s,m}=m_sJ_m$ and assign the remaining background energy to the visible bath, $\rho_{v,m}=3M_{\rm P}^2H_m^2-\rho_{s,m}$, with $H_m=H(N_m)$. We set $\rho_{a,\rm rel,m}=0$ because this component counts axions produced by subsequent saxion decay. The angular waves generated during relaxation form a separate component whose small energy density, quoted above, is neglected in this matching.

The visible bath contains $g_{\rm SM}=106.75$ relativistic Standard Model degrees of freedom and equilibrium KSVZ quarks with $N_c=3$. We follow their disappearance below $T\sim m_\psi$ using the Fermi--Dirac distribution at zero chemical potential~\cite{Laine:2016hma}. Define $z=m_\psi/T$, $q=p/T$, $e_q=\sqrt{q^2+z^2}$ and $n_F(e_q)=(e^{e_q}+1)^{-1}$, where $p$ is the physical quark momentum. The visible energy density and pressure are
\begin{equation}
 \begin{aligned}
 \rho_v(T)&=\frac{\pi^2g_{\rm SM}}{30}T^4
 +\frac{2N_cT^4}{\pi^2}\int_0^\infty dq\,q^2e_qn_F(e_q),\\
 p_v(T)&=\frac{\pi^2g_{\rm SM}}{90}T^4
 +\frac{2N_cT^4}{3\pi^2}\int_0^\infty dq\,\frac{q^4}{e_q}n_F(e_q).
 \end{aligned}
 \label{eq:app_late_visible_EOS}
\end{equation}
The quark integrals include spin, antiparticle and colour states. The effective energy degrees of freedom decrease from 117.25 to 106.75 as these quarks disappear.

Saxion decay into KSVZ quarks, followed by their rapid decay into Standard Model particles, heats the visible bath and increases its entropy~\cite{Lyth:1993zw}. We evolve its temperature using $C_v=d\rho_v/d\ln T$, the energy-density response to temperature, and $s_v=(\rho_v+p_v)/T$, the entropy density at zero chemical potential~\cite{Laine:2016hma}. The visible continuity equation gives
\begin{equation}
 \begin{aligned}
 \frac{d\ln T}{d\ln a}&=
 \frac{-3(\rho_v+p_v)+(\Gamma^{(0)}_{\psi\bar\psi}/H)\rho_s}{C_v},\\
 \frac{d(a^3s_v)}{dt}&=
 \frac{a^3\Gamma^{(0)}_{\psi\bar\psi}\rho_s}{T}.
 \end{aligned}
 \label{eq:app_late_temperature_ode}
\end{equation}
These equations include expansion cooling, decay heating and the changing quark population. We solve them with Eqs.~\eqref{eq:late_saxion_fluid}--\eqref{eq:late_Friedmann} until $\Gamma_s/H=100$, when the condensate has effectively decayed. The visible comoving entropy $S=a^3s_v$ increases by $S_f/S_i=1.0049$ in A and $1.0186$ in B, where $i$ and $f$ denote the initial and final fluid epochs. The corresponding correction to Eq.~\eqref{eq:num_reference_matching} is $-\ln(S_f/S_i)/3$, with magnitude below $0.0062$ e-fold.

Residual radial oscillations can also amplify axions through parametric resonance~\cite{Kofman:1997yn,Ema:2017krp}. For $\overline X=f_a[1+\varepsilon\cos(m_st)]$, the relative amplitude $\varepsilon\ll1$ varies slowly over an oscillation; reflection gives the motion about the negative minimum. The leading narrow resonance band is centred at $p=m_s/2$, and expansion limits the time each mode spends in the band~\cite{Moroi:2020bkq}. Treating $H$ and $\varepsilon$ as constant during one crossing gives the occupation $n_k=e^C-1$ for a mode entering in the vacuum. The growth exponent $C$ and the axion energy injected per e-fold are
\begin{equation}
 C=\frac{\pi\varepsilon^2m_s}{2H},
 \qquad
 \left.\frac{d\rho_{a,\rm rel}}{dN}\right|_{\rm source}
 =\frac{m_s^4}{32\pi^2}(e^C-1).
 \label{eq:revised_narrow_occupation}
\end{equation}
The source is evaluated before subsequent redshifting. With $\rho_s\simeq m_s^2f_a^2\varepsilon^2/2$, its term linear in $C$ equals $\Gamma_{aa}^{(0)}\rho_s/H$, the perturbative decay source already present in the fluid equations. We therefore retain only the additional contribution proportional to $e^C-1-C$. During radiation domination with fixed relativistic degrees of freedom, $\varepsilon^2\propto a^{-3}\exp[-\int\Gamma_sdt]$ and $H\propto a^{-2}$, giving $C\propto a^{-1}\exp[-\int\Gamma_sdt]$. The additional production thus weakens as the condensate dilutes and decays.

We check this estimate by evolving five transverse modes in each of benchmarks A and B, spanning approximately $15\lesssim\kappa\lesssim28$. Their final occupations agree with the narrow-band result to better than $0.2\%$. Integrating the additional source gives fractional condensate depletions of $3.1\times10^{-5}$ in A and $3.8\times10^{-5}$ in B. Within this approximation, allowing for the small departure from radiation domination keeps the correction below roughly $10^{-4}$.

Each axion from a nonrelativistic saxion decay has momentum approximately $m_s/2$, which subsequently redshifts as $a^{-1}$. For a representative decay at $T_\Gamma=T(H=\Gamma_s)$, neglecting the small remaining entropy injection gives the present momentum
$p_{a,0}\simeq(m_s/2)(T_{\gamma,0}/T_\Gamma)(g_{*s,0}/g_{*s,d})^{1/3}\sim10^{-3}\,\mathrm{eV}$.
Here $g_{*s,d}=106.75$ is the visible entropy count after decay, while $g_{*s,0}$ and $T_{\gamma,0}$ are defined in Appendix~\ref{app:background_methods}. This momentum exceeds the vacuum axion mass, approximately $6\times10^{-8}\,\mathrm{eV}$ at $f_a=10^{14}\,\mathrm{GeV}$~\cite{GrillidiCortona:2015jxo,Gorghetto:2018ocs}, so the axion population carrying the decay radiation remains relativistic today.

Finally, the additional radiation raises $H$ at fixed visible temperature and delays the onset of cold axion misalignment oscillations to a lower temperature~\cite{Arias:2021rer}. We define the onset temperature $T_{\rm osc}$ by $m_a(T_{\rm osc})\simeq3H(T_{\rm osc})$, where $m_a(T)$ is the temperature-dependent mass determined by the QCD topological susceptibility~\cite{Borsanyi:2016ksw,Petreczky:2016vrs,Bonati:2015vqz,Chen:2022fid,Athenodorou:2022aay,Bonanno:2025wcv}. Over the onset interval, we approximate $m_a(T)\propto T^{-n}$, with local exponent $n$, and define $R_{\rm DR}=\rho_{a,\rm rel}/\rho_v$. Holding this ratio and the relativistic degrees of freedom constant gives $H(T)\propto T^2\sqrt{1+R_{\rm DR}}$ and hence $T_{\rm osc}\propto(1+R_{\rm DR})^{-1/[2(n+2)]}$.

The harmonic axion number density at onset is proportional to $m_a(T_{\rm osc})f_a^2\theta_i^2$. At fixed $f_a$ and $\theta_i$, conservation of axion number per visible entropy therefore gives a present abundance proportional to $m_a(T_{\rm osc})/T_{\rm osc}^3$. Combining these relations yields
\begin{equation}
 \frac{\Omega_a(R_{\rm DR})}{\Omega_a(0)}
 \simeq(1+R_{\rm DR})^{(n+3)/[2(n+2)]},
 \label{eq:misalignment_extra_radiation}
\end{equation}
where $\Omega_a(0)$ is the abundance without additional radiation at the same initial angle. For $n=4$ and $g_{*s}=60$ at onset, the abundance increases by approximately $0.6\%$ in A and $2\%$ in B. Since the harmonic abundance scales as $\theta_i^2$, retaining axions as all dark matter requires angles smaller by $0.3\%$ and $1\%$, respectively. These shifts quantify the correction to the common misalignment angle in Table~\ref{tab:cosmological_benchmarks}.

\FloatBarrier

\bibliographystyle{JHEP}
\bibliography{Referencxes}
\end{document}